\documentclass[twocolumn]{aastex62}

\graphicspath{{./}}

\accepted{\today}
\shorttitle{ASPECS: The cosmic dust and gas mass densities in galaxies}
\shortauthors{Magnelli et al.}
\begin{document}

\title{The ALMA Spectroscopic Survey in the HUDF: The Cosmic Dust and Gas Mass Densities in Galaxies up to $z\sim3$}

\correspondingauthor{Benjamin Magnelli}
\email{magnelli@astro.uni-bonn.de}

\author[0000-0002-6777-6490]{Benjamin Magnelli}
\affiliation{Argelander Institut f\"ur Astronomie, Universit\"at Bonn, Auf dem H\"ugel 71, Bonn, D-53121, Germany}

%%%% 1st TIER

\author[0000-0002-3952-8588]{Leindert Boogaard}
\affil{Leiden Observatory, Leiden University, PO Box 9513, NL-2300 RA Leiden, The Netherlands}

\author[0000-0002-2662-8803]{Roberto Decarli}
\affiliation{INAF-Osservatorio di Astrofisica e Scienza dello Spazio, via Gobetti 93/3, I-40129, Bologna, Italy}

\author[0000-0003-3926-1411]{Jorge G\'onzalez-L\'opez}
\affil{Las Campanas Observatory, Carnegie Institution of Washington, Casilla 601, La Serena, Chile}
\affil{N\'ucleo de Astronom\'ia de la Facultad de Ingenier\'ia y Ciencias, Universidad Diego Portales, Av. Ej\'ercito Libertador 441, Santiago, Chile}

\author[0000-0001-8695-825X]{Mladen Novak}
\affil{Max Planck Institut f\"ur Astronomie, K\"onigstuhl 17, 69117 Heidelberg, Germany}

\author[0000-0003-1151-4659]{Gerg\"o Popping}
\affil{Max Planck Institut f\"ur Astronomie, K\"onigstuhl 17, 69117 Heidelberg, Germany}
\affil{European Southern Observatory, Karl Schwarzschild Strasse 2, 85748 Garching, Germany}

\author[0000-0003-3037-257X]{Ian Smail}
\affil{Centre for Extragalactic Astronomy, Department of Physics, Durham University, South Road, Durham, DH1 3LE, UK}

\author[0000-0003-4793-7880]{Fabian Walter}
\affiliation{Max Planck Institut f\"ur Astronomie, K\"onigstuhl 17, 69117 Heidelberg, Germany}
\affiliation{National Radio Astronomy Observatory, Pete V. Domenici Array Science Center, P.O. Box O, Socorro, NM 87801, USA}

%%%% 2nd TIER

\author[0000-0002-6290-3198]{Manuel Aravena}
\affil{N\'ucleo de Astronom\'ia, Facultad de Ingenier\'ia y Ciencias, Universidad Diego Portales, Av. Ej\'ercito 441, Santiago, Chile}

\author[0000-0002-9508-3667]{Roberto J. Assef}
\affil{N\'ucleo de Astronom\'ia, Facultad de Ingenier\'ia y Ciencias, Universidad Diego Portales, Av. Ej\'ercito 441, Santiago, Chile}

\author[0000-0002-8686-8737]{Franz Erik Bauer}
\affil{Instituto de Astrof\'isica, Facultad de F\'isica, Pontificia Universidad Cat\'olica de Chile Av. Vicu\~na Mackenna 4860, 782-0436 Macul, Santiago, Chile}
\affil{Millennium Institute of Astrophysics (MAS), Nuncio Monse\~nor S\'otero Sanz 100, Providencia, Santiago, Chile}
\affil{Space Science Institute, 4750 Walnut Street, Suite 205, Boulder, CO 80301, USA}

\author[0000-0002-1707-1775]{Frank Bertoldi}
\affil{Argelander Institut f\"ur Astronomie, Universit\"at Bonn, Auf dem H\"ugel 71, Bonn, D-53121, Germany}

\author[0000-0001-6647-3861]{Chris Carilli}
\affil{National Radio Astronomy Observatory, Pete V. Domenici Array Science Center, P.O. Box O, Socorro, NM 87801, USA}
\affil{Battcock Centre for Experimental Astrophysics, Cavendish Laboratory, Cambridge CB3 0HE, UK}

\author{Paulo C. Cortes}
\affil{Joint ALMA Observatory - ESO, Av. Alonso de C\'ordova, 3104, Santiago, Chile}
\affil{National Radio Astronomy Observatory, 520 Edgemont Rd, Charlottesville, VA, 22903, USA}

\author[0000-0001-9759-4797]{Elisabete da Cunha}
\affil{International Centre for Radio Astronomy Research, The University of Western Australia, 35 Stirling Highway, Crawley, WA 6009, Australia}

\author[0000-0001-6647-3861]{Emanuele Daddi}
\affil{Laboratoire AIM, CEA/DSM-CNRS-Universit\'e Paris Diderot, Irfu/Service d'Astrophysique, CEA Saclay, Orme des Merisiers, 91191 Gif-sur-Yvette cedex, France}

\author[0000-0003-0699-6083]{Tanio D\'iaz-Santos}
\affil{N\'ucleo de Astronom\'ia, Facultad de Ingenier\'ia y Ciencias, Universidad Diego Portales, Av. Ej\'ercito 441, Santiago, Chile}

\author[0000-0003-4268-0393]{Hanae Inami}
\affil{Hiroshima Astrophysical Science Center, Hiroshima University, 1-3-1 Kagamiyama, Higashi-Hiroshima, Hiroshima, 739-8526, Japan}

\author[0000-0001-5118-1313]{Robert J. Ivison}
\affil{European Southern Observatory, Karl Schwarzschild Strasse 2, 85748 Garching, Germany}
\affil{Institute for Astronomy, University of Edinburgh, Royal Observatory, Blackford Hill, Edinburgh EH9 3HJ, UK}

\author{Olivier Le F\`evre}
\affil{Aix Marseille Universit\'e, CNRS, LAM (Laboratoire d'Astrophysique de Marseille) UMR 7326, 13388, Marseille, France}

\author[0000-0001-5851-6649]{Pascal Oesch}
\affil{Department of Astronomy, University of Geneva, 51 Ch. des Maillettes, 1290 Versoix, Switzerland}
\affil{International Associate, Cosmic Dawn Center (DAWN) at the Niels Bohr Institute, University of Copenhagen and DTU-Space, Technical University of Denmark, Copenhagen, Denmark}

\author[0000-0001- 9585-1462]{Dominik Riechers}
\affil{Department of Astronomy, Cornell University, Space Sciences Building, Ithaca, NY 14853, USA}
\affil{Max Planck Institut f\"ur Astronomie, K\"onigstuhl 17, 69117 Heidelberg, Germany}
\affil{Humboldt Research Fellow}

\author[0000-0003-4996-9069]{Hans-Walter Rix}
\affiliation{Max Planck Institut f\"ur Astronomie, K\"onigstuhl 17, 69117 Heidelberg, Germany}

\author[0000-0003-1033-9684]{Mark T. Sargent}
\affil{Astronomy Centre, Department of Physics and Astronomy, University of Sussex, Brighton, BN1 9QH, UK}

\author[0000-0001-5434-5942]{Paul van der Werf}
\affil{Leiden Observatory, Leiden University, PO Box 9513, NL-2300 RA Leiden, The Netherlands}

\author{Jeff Wagg}
\affil{SKA Organization, Lower Withington Macclesfield, Cheshire SK11 9DL, UK}

\author{Axel Weiss}
\affil{Max-Planck-Institut f\"ur Radioastronomie, Auf dem H\"ugel 69, 53121 Bonn, Germany}

\begin{abstract}
Using the deepest 1.2\,mm continuum map to date in the Hubble Ultra Deep Field obtained as part of the ALMA Spectroscopic Survey (ASPECS) large program, we measure the cosmic density of dust and implied gas (H$_{2}+$\ion{H}{1}) mass in galaxies as a function of look--back time. We do so by stacking the contribution from all $H$-band selected galaxies above a given stellar mass in distinct redshift bins, $\rho_{\rm dust}(M_\ast>M,z)$ and $\rho_{\rm gas}(M_\ast>M,z)$. At all redshifts, $\rho_{\rm dust}(M_\ast>M,z)$ and $\rho_{\rm gas}(M_\ast>M,z)$ grow rapidly as $M$ decreases down to $10^{10}\,M_\odot$, but this growth slows down towards lower stellar masses. This flattening implies that at our stellar mass-completeness limits ($10^8\,M_\odot$ and $10^{8.9}\,M_\odot$ at $z\sim0.4$ and $z\sim3$), both quantities converge towards the total cosmic dust and gas mass densities in galaxies. The cosmic dust and gas mass densities increase at early cosmic time, peak around $z\sim2$, and decrease by a factor $\sim4$ and 7, compared to the density of dust and molecular gas in the local universe, respectively. The contribution of quiescent galaxies -- i.e., with little on-going star-formation-- to the cosmic dust and gas mass densities is minor ($\lesssim10\%$). The redshift evolution of the cosmic gas mass density resembles that of the star-formation rate density, as previously found by CO-based measurements. This confirms that galaxies have relatively constant star-formation efficiencies (within a factor $\sim2$) across cosmic time. Our results also imply that by $z\sim0$, a large fraction ($\sim90\%$) of dust formed in galaxies across cosmic time has been destroyed or ejected to the intergalactic medium.
\end{abstract}

\keywords{galaxies: evolution -- galaxies: formation -- galaxies: high redshift}

\section{Introduction}
\label{sec:intro}
The cosmic star formation rate density (SFRD) of the Universe (i.e., mass of stars formed per unit time and comoving volume; $\rho_{\rm SFR}$) evolves significantly with redshift \citep[see][for a review]{madau_2014}.
It increased from early cosmic epochs, peaked at $z=1-3$, and then decreased steadily until the present day.
To understand this evolution and therefore how galaxies formed and evolved throughout cosmic time, it is necessary to study their molecular gas reservoirs --i.e., the phase out of which stars form-- and measure the evolution of the cosmic molecular gas mass density.
There are different approaches to measuring these gas reservoirs, the fundamental problem being that molecular hydrogen (H$_2$, the main constituent of the molecular gas) cannot be observed easily at the mass-weighted temperatures and density of the cold star--forming interstellar medium (ISM).

Traditionally, the emission lines of the different rotational states of the carbon monoxide molecule ($^{12}$CO; hereafter, CO) have been used as tracer of the molecular gas \citep[see][for a review]{bolatto_2013}, but there are other tracers as well.
Most notably, the continuum emission from dust is frequently used as an alternative tracer of the gas, though including both the molecular (H$_{2}$) and atomic (\ion{H}{1}) phases. 
With the advent of the \textit{Herschel} Space Observatory, such dust-based gas mass estimates have been used for high-redshift galaxies \citep[e.g.,][]{magdis_2012,magnelli_2012,santini_2014,genzel_2015}, by fitting their far-infrared-to-submillimeter emission with dust models \citep[e.g.,][]{draine_2007} and using the local gas-to-dust mass ratio relation \citep[e.g.,][]{leroy_2011}.
Most recently, \citet{scoville_2014,scoville_2016,scoville_2017} advocated that accurate dust-based gas mass estimates could be inferred using a single dust emission measurement in the Rayleigh-Jeans tail.
This method relies on the assumption that the mass-dominant dust component of the ISM of most galaxies is at around 25\,K, that this component accounts for the bulk of their Rayleigh-Jeans emission, and that the emission is optically thin (see Section~\ref{subsec:measuring_mdust}).
With this approach, \citet{scoville_2016} calibrated a single conversion factor ($\alpha_{\rm 850\,\mu m}$) from the Rayleigh-Jeans dust emission to the (molecular) gas mass of massive galaxies ($>10^{10}\,M_\odot$).
Interestingly, this conversion factor is consistent within a few percent of that inferred from the dust mass absorption cross section of \citet{draine_2007} and the typical gas-to-dust mass ratio of 100 for massive galaxies at $z\sim0$ \citep{leroy_2011}. 
Dust-based gas mass estimates from far-infrared-to-submillimeter fits using the \citet{draine_2007} model and from the Rayleigh-Jeans method of \citet{scoville_2016} are thus consistent for massive galaxies with a typical gas-to-dust mass ratio of 100 \citep[e.g.,][]{magnelli_2019}.

In recent years, using both methods, numerous studies have reached a common conclusion: at high-redshift ($z\gtrsim0.2$) dust-based gas mass estimates are consistent within $\sim0.2\,$dex with those inferred from CO line emission \citep{genzel_2015,scoville_2016,scoville_2017,tacconi_2018,kaasinen_2019}.
This has demonstrated the reliability of dust-based gas mass measurements for massive ($>10^{10}\,M_\odot$) galaxies and suggested that the gas in these galaxies is mostly dominated by its molecular phase.
It has thereby facilitated the study of the gas content in high-redshift massive galaxies, which can be measured in just a few minutes of observing time with the Atacama Large Millimeter Array (ALMA).

From these dust-based gas mass estimates and the ever growing sample of CO measurements, much has been learned in recent years about the gas reservoirs of massive high-redshift galaxies \citep[e.g.,][]{magdis_2012,magnelli_2012,bothwell_2013,saintonge_2013,santini_2014,genzel_2015,scoville_2016,schinnerer_2016,aravena_2016,decarli_2016b,scoville_2017,tacconi_2018,kaasinen_2019}.
It is now robustly established that the gas fraction, $f_{\rm gas}=M_{\rm gas}/M_\ast$, in massive galaxies steadily decreases between $z\sim4$ and $z\sim0$, while their star formation efficiency (i.e., SFR$/M_{\rm gas}$) only slightly decreases within this redshift range.
Larger gas supply rather than enhanced star-formation efficiency, seems thus to explain the elevated specific star formation rate (SSFR; SFR$/M_{\ast}$) of massive high-redshift galaxies compared to galaxies in the local Universe \citep[e.g.,][]{schreiber_2015}.

While these results are of utmost importance for galaxy evolution models, they can, however, not easily be extrapolated to infer the redshift evolution of the cosmic gas mass density in galaxies.
Indeed, these targeted studies are biased towards massive, star-forming galaxies and could thus miss a significant fraction of gas-rich galaxies in the Universe.
Blind spectroscopic surveys at millimeter and radio wavelengths provide here a complementary approach.
The ALMA Spectroscopic Survey pilot and large program \citep[ASPECS pilot and ASPECS LP, respectively;][]{walter_2016,decarli_2016,decarli_2019} as well as the Jansky Very Large Array COLDz survey \citep{riechers_2019} have in particular been used to constrain the CO luminosity function and thereby the cosmic molecular gas mass density from $z\sim4$ to $z\sim0.3$.
These studies revealed that the cosmic molecular gas mass density closely matches the evolution of the cosmic SFRD, implying that the average star formation efficiency in galaxies did not significantly evolve with redshift.
Naturally, these studies also suffer from a number of limitations and in particular their dependencies on the assumed CO excitation and CO-to-H$_{2}$ conversion factors.
To alleviate some of these uncertainties, independent constraints on the cosmic gas mass density using dust-based gas mass estimates are needed.
Such studies would simultaneously measure the redshift evolution of the cosmic dust mass density in galaxies, which to date remains only sparsely constrained \citep[e.g.,][]{dunne_2003,dunne_2011,driver_2018,pozzi_2019}.
This latter measurement would be instrumental for the growing number of galaxy evolution models that track self-consistently the production and destruction of dust \citep[e.g.,][]{popping_2017,aoyama_2018,vijayan_2019,dave_2019}. 

As part of the ASPECS LP, we obtained the deepest ALMA $1.2$\,mm continuum map of the \textit{Hubble} Ultradeep Field \citep[HUDF; 1$\sigma$ sensitivity of $9.5\,\mu$Jy\,beam$^{-1}$; area of 4.2\,arcmin$^2$;][Gonzalez--Lopez et al.\ 2019b]{walter_2016}.
The wavelength of this survey probes the Rayleigh-Jeans dust emission of galaxies up to $z\sim4$ and is thus ideal to measure the dust and implied gas mass of high-redshift galaxies using the method advocated by \citet{scoville_2016}.
In addition, because this is a blind survey, it provides an unbiased view on the observed-frame 1.2\,mm emission from all galaxies\footnote{the only exception being galaxies with very extended emission ($\gtrsim\,10\arcsec$), which could be missed by our observations due to the lack of very short baselines. However, such extended emission would only be associated to low-redshift galaxies ($z\lesssim0.1$) which are not studied in the present paper.} within the comoving volume probed by our map.
While a fraction of this 1.2\,mm emission is included in individual detections, a large portion could, however, reside below our detection threshold, even in the case of this deep 1.2\,mm map.
Fortunately, the HUDF is one of the best studied extragalactic regions in the sky.
It thus benefits from a remarkable wealth of ancillary data, providing a unique opportunity to build stellar mass--complete sample of galaxies down to, e.g., $\sim\,10^{8}\,M_\odot$ and $\sim\,10^{8.9}\,M_\odot$ at $z\sim0.4$ and 3, respectively \citep[e.g.,][]{mortlock_2015}.
Knowing \textit{a priori} the positions of this stellar mass-complete sample of galaxies, we can thus sum up their 1.2\,mm emission within a given comoving volume, irrespective of their individual detectability in our 1.2\,mm map.
Converting this 1.2\,mm emission per unit comoving volume into dust and gas masses, assuming $T_{\rm dust}=25\,$K \citep{scoville_2016} and the local gas-to-dust mass ratio relation \citep{leroy_2011}, we can measure the cosmic dust and gas mass densities of all the known galaxies in the HUDF above a given stellar mass and as a function of look--back time, i.e., $\rho_{\rm dust}(M_\ast>M,z)$ and $\rho_{\rm gas}(M_\ast>M,z)$.
These measurements provide constraints for galaxy evolution models and complement those inferred from the ASPECS CO survey \citep{decarli_2016,decarli_2019}. 

The structure of this paper is as follows. 
In Section~\ref{sec:data}, we present the ASPECS LP 1.2\,mm continuum map.
In Section~\ref{sec:sample}, we summarize the ancillary data used in this study and the build-up of our stellar mass-complete sample of galaxies.
In Section~\ref{sec:methodology}, we present the method used to measure the cosmic dust and gas mass densities through stacking of the ASPECS LP 1.2\,mm map.
In Section~\ref{sec:rho_dust} and ~\ref{sec:rho_gas}, we present the main results of this study, i.e., the evolution of cosmic dust and gas mass densities as a function of stellar mass and look--back time.
In Section~\ref{sec:discussion}, we compare these results with outputs from simulations and discuss implications for galaxy evolution models.
Finally, in Section~\ref{sec:conclusion} we present our conclusions.

Throughout this paper, we assume a $\Lambda$CDM cosmology, adopting $H_{0}=67.8$~(km/s)/Mpc, $\Omega_{\rm M}=0.308$ and $\Omega_{\Lambda}=0.692$ \citep{planck_cosmo_2016}.
At $z=1$, 1$\arcsec$ corresponds to 8.229\,kpc.
A \citet{chabrier_2003} initial mass function (IMF) is used for all stellar masses quoted in this article.

\section{Data}
\label{sec:data}
The ASPECS LP 1.2\,mm survey covers 4.2~arcmin$^2$ in the HUDF, centered at $\alpha$ $=$ 3$^{\rm h}$ 32$^{\rm m}$ 38.5$^{\rm s}$ , $\delta$ $=$ -27$^\circ$ 47$\arcmin$ 00$\arcsec$ (J2000; 2016.1.00324.L).
The survey strategy as well as the data calibration and imaging are described in detail by Gonzalez--Lopez et al.\ (2019b).
Here we only summarize the most important information.

The ASPECS LP 1.2\,mm continuum map was obtained by combining eight spectral tunings that cover most of the ALMA band 6.
The mosaic consists of 85 pointings and is Nyquist-sampled at all wavelengths.
The data was calibrated with the Common Astronomy Software Applications \citep[\textsc{CASA};][]{mcMullin_2007} calibration pipeline using the script provided by the Joint ALMA Observatory (JAO).
Imaging was also done in \textsc{CASA} using the multi-frequency synthesis algorithm implemented within the task \textsc{TCLEAN}, which combines all pointings together down to a primary beam (PB) gain of 0.1.
We used natural weighting and `cleaned' down to 20\,$\mu$Jy\,beam$^{-1}$ all sources with a signal-to-noise ratio (SNR) greater than 5.
The synthesized beam has a full width at half maximum (FWHM) of $1\farcs53\times1\farcs08$ at a position angle of $-84^\circ$.
The mosaic covers 2.9 and 4.2 arcmin$^2$ of the HUDF, within a combined PB coverage\footnote{this corresponds to the `.pb' array output by the task \textsc{TCLEAN} when used in `mosaic' mode.} of 50\% and 10\%, respectively.
The deepest region in the map (i.e., with a combined PB coverage $\gtrsim90\%$) has an rms of $9.5\,\mu$Jy\,beam$^{-1}$.
Where the combined PB coverage is better than 75\% (i.e., the region of interest of our study; see Section~\ref{subsec:stacking}), we detected 22 galaxies with a SNR greater than 3 and a `Fidelity' factor greater than 0.5.

\section{Sample}
\label{sec:sample}
\begin{figure*}
\begin{center}
\includegraphics[width=\linewidth]{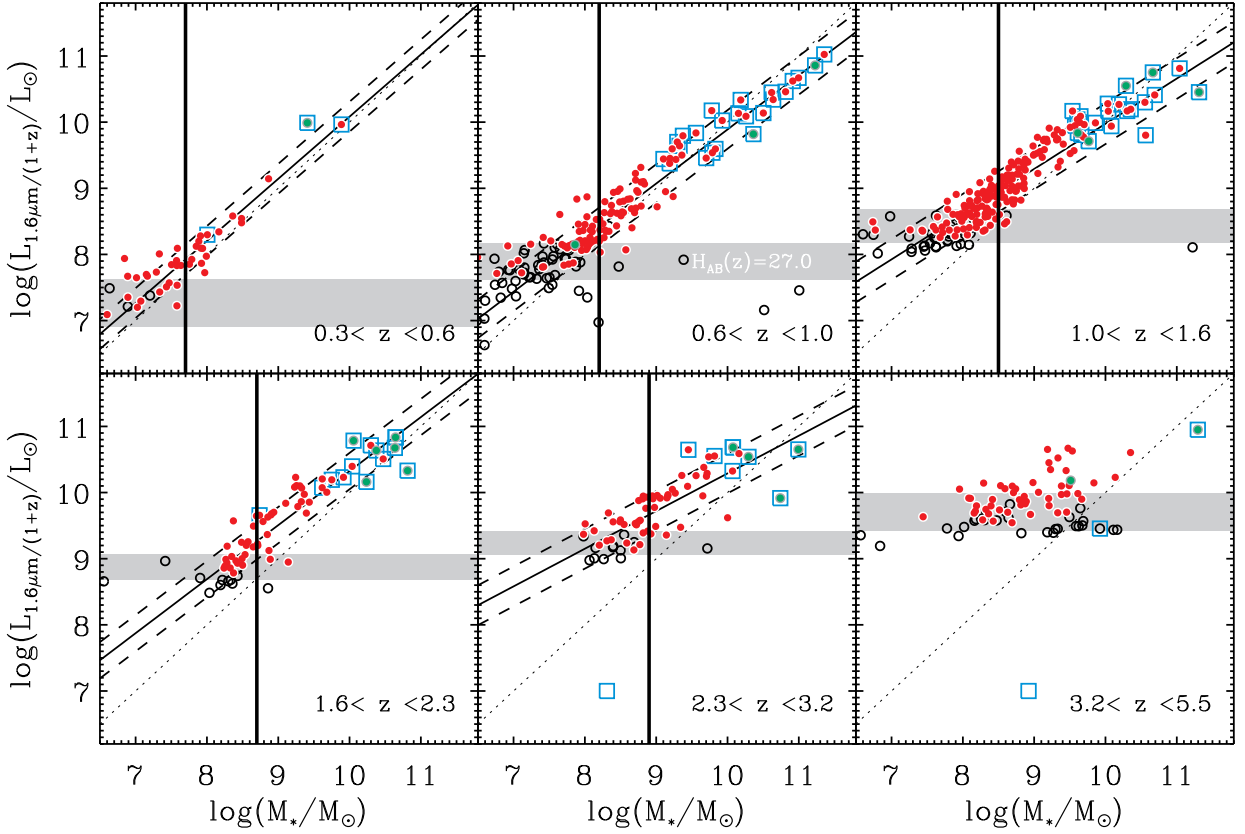}	
\caption{\label{fig:lh_vs_mass}
        Correlation in different redshift bins between the stellar mass and observed-frame $H$-band luminosity of our $H$-band-selected galaxies (i.e., $H$$\,\leq\,$$27$; filled circles; 555 galaxies; green filled circles represent galaxies individually detected at 1.2\,mm --21 galaxies--, while red filled circles are galaxies undetected at 1.2\,mm --534 galaxies).
        Open circles present galaxies within the XDF but with $H$$\,>\,$$27$, blue squares show galaxies detected by IRAC at $5.8\,\mu$m.
        For galaxies detected by IRAC at $5.8\,\mu$m but not detected in the $H$-band (2 galaxies), we artificially set their observed-frame $H$-band luminosity to $10^7\,L_\odot$. 
        The shaded areas define the range of our $H$$\,=\,$$27$ selection limit at the lowest and highest redshift of each bin.
        The black lines are the best-fit relations, while the dashed lines show their 1$\,\sigma$ dispersion.
        The thick black vertical lines correspond to the stellar mass-completeness limits of our $H$-band-selected sample, defined at the intersection between our $H$-band selection limits at the highest redshift of the bin (upper horizontal boundary of the gray shaded area) and the lower l$\,\sigma$ boundary of the best-fit relation.
        The thin dotted lines show the one-to-one relation.
        At $z\gtrsim3.2$, the observed-frame $H$-band luminosities of galaxies do not strongly correlate with their stellar masses, as expected (see text).
        Thus, at these redshifts, our $H$-band-selected sample cannot be considered as stellar mass--selected and therefore stellar mass--complete.
		}
\end{center}
\vspace{0.2cm}
\end{figure*}

The ASPECS LP 1.2\,mm survey covers most of the \textit{Hubble} eXtremely Deep Field \citep[XDF;][]{illingworth_2013,koekemoer_2013}, itself located within the HUDF \citep[][]{beckwith_2006}.
It is one of the best studied extragalactic regions in the sky, and thereby benefits from a remarkable wealth of ancillary data. 
The compilation of our master catalogue of galaxies and the modelling of their spectral energy distribution (SED) are described by \citet{decarli_2019} and \citet{boogaard_2019}, respectively. 
Here, we summarize the most important information.

In the XDF, the bulk of the optical and near-infrared observations comes from the \textit{Hubble Space Telescope} (\textit{HST}) as part of the Cosmic Assembly Near-infrared Deep Extragalactic Legacy Survey \citep[CANDELS;][]{grogin_2011,koekemoer_2011} and the HUDF09 \citep[e.g.,][]{bouwens_2011} and HUDF12 \citep{koekemoer_2013} surveys. 
These observations were obtained with the Advanced Camera for Surveys (ACS) at optical wavelengths and with the Wide Field Camera 3 (WFC3) in the near-infrared (NIR).
By combining all the \textit{HST} observations, \citet{skelton_2014} performed a multi-wavelength photometric analysis, which also included publicly available ground based optical/NIR \citep[see][and reference therein]{skelton_2014} as well as \textit{Spitzer}-IRAC images \citep{labbe_2015}.
Complemented with \textit{Spitzer} MIPS--24\,$\mu$m photometry from \citet{whitaker_2014}, this constitutes our master photometric catalogue.
It provides measurements in $>\,$30 broad and medium bands for 1481 sources in the region of interest in the XDF, i.e., where the combined PB coverage of the ASPECS LP 1.2\,mm survey is better than $75\%$ (see Section~\ref{subsec:stacking}).

The spectroscopic redshifts of 443 of these galaxies were obtained from a variety of studies: the MUSE \textit{Hubble} Ultra Deep Survey \citep{bacon_2017,inami_2017}; \textit{HST} grism spectroscopy in the optical \citep{xu_2007} and in the NIR \citep[3D-\textit{HST} survey;][]{moncheva_2016}; and spectroscopic compilations from \citet{lefevre_2005}, \citet{Coe_2006}, \citet{skelton_2014}, and \citet{morris_2015}.
For galaxies with no spectroscopic redshift available, we use photometric redshifts determined in \citet{skelton_2014} by means of the \texttt{EAZY} code, with a typical uncertainty, $\sigma[\,|z_{\rm phot}-z_{\rm spec}|/(1+z_{\rm spec})\,]$, of 0.010, and only $5.4$\% of objects with $|z_{\rm phot}-z_{\rm spec}|/(1+z_{\rm spec})\,>0.1$. 

The stellar mass of each galaxy was obtained by modelling their SED, using the high-redshift extension of the \texttt{MAGPHYS} code \citep{dacunha_2008,dacunha_2015}.
We used the galaxy's photometry between 0.37$\,\mu$m and 8.0$\,\mu$m, as well as their 1.2\,mm flux density (or upper limit).
These stellar masses are on average $\sim0.2$\,dex larger than those measured by \citet{skelton_2014} using the \texttt{FAST} code.
We verified that our results on the cosmic dust and gas mass densities remain unchanged --simply shifted towards lower stellar masses--, while using the stellar masses of \citet{skelton_2014}.

From this master catalogue, we kept only the 555 galaxies with an observed total\footnote{we refer the reader to \citet{skelton_2014} for details on how these `total' $H$-band magnitude were measured.} $H$-band magnitude brighter than 27, of which 281 had a spectroscopic redshift.
This NIR selection was chosen because the observed-frame $H$-band luminosity of a galaxy is known to correlate with its stellar mass up to $z\sim3$ (Figure~\ref{fig:lh_vs_mass}).
Furthermore, in our master catalogue, the $H$-band magnitude distribution (i.e., number of galaxies per bin of magnitude) peaks at $H=27$ and rapidly decreases towards fainter magnitudes.
This rapid decrease at $H$$\,>\,$$27$ suggests that at such faint magnitudes, our catalogue is affected by large photometric incompleteness.
We thus restricted our analysis to sources with $H$$\,\leq\,$$27$.   
This $H\leq27$ selection also ensured that the number of broad and medium bands available for each source was high enough ($15^{+8}_{-7}$, median and 16th and 84th percentiles) for accurate SED modelling and thus stellar mass estimates. 

As a last step, we determined down to which stellar masses our $H$-band-selected galaxy sample can be considered stellar mass--complete.
To this end, we used the empirical approach described in \citet{schreiber_2015}.
For each redshift bin of interest, we fitted the correlation between the observed-frame $H$--band (1.6\,$\mu$m) luminosities of galaxies and their stellar masses (derived as described above) with a simple power law, i.e., $M_{\ast}$$\,=\,$$CL^{\alpha}$ (Figure~\ref{fig:lh_vs_mass}).
Then, we estimated the scatter around this correlation, which is caused by differences of age, attenuation and $k$-correction between these galaxies.
Finally, for a given redshift bin, the stellar mass-completeness limit was set by the stellar mass corresponding to the $H$--band luminosity cut plus the 1$\,\sigma$ dispersion of the $L_{1.6\,\mu{\rm m}/(1+z)}-M_{\ast}$ relation.   
At this stellar mass-completeness limit, only 16\% of galaxies are expected to be missed because of our $H\leq27$ selection criterion, while this percentage drop rapidly to 0\% towards higher stellar masses (see open symbols in Figure~\ref{fig:lh_vs_mass}).
At $z\gtrsim3.2$, because the $H$-band probes rest-wavelengths shorter than 4000$\,$\AA\ (i.e., Balmer break), the observed-frame $H$-band luminosity of a galaxy does not anymore correlate strongly with its stellar mass (Figure~\ref{fig:lh_vs_mass}).
At these redshifts, our $H$-band-selected sample cannot be considered as stellar mass--selected and therefore stellar mass--complete.

We note that while the observed-frame IRAC luminosities of galaxies correlate in principle better with their stellar masses than the observed-frame $H$-band luminosities, IRAC observations in the XDF do not provide stellar mass-complete samples as deep as that provided by the $H$-band.
To illustrate this, we plot in Figure~\ref{fig:lh_vs_mass} the locus of all galaxies detected by IRAC at $5.8\,\mu$m within our region of interest in the XDF, including those not detected in the $H$-band (2 galaxies) and to which we artificially attributed an observed-frame $H$-band luminosity of $10^7\,L_\odot$.
IRAC $5.8\,\mu$m-selected galaxies are associated to the brightest and most massive galaxies in our $H$-band-selected catalogue.
At $z<3$, only one IRAC $5.8\,\mu$m-selected galaxies is missed by our $H$-band selection, and it has a stellar mass below our stellar mass completeness limit. 
This supports the assumption that the observed-frame $H$--band luminosity of a galaxy is a good proxy of its stellar mass up to $z<3$, and that in the XDF, a $H$-band-selected catalogue has a much lower stellar mass-completeness limits than an IRAC $5.8\,\mu$m-selected catalogue.
Repeating this analysis with IRAC $3.6\,\mu$m leads to the same conclusions, though with stellar mass-completeness limits getting closer to that of our $H$-band-selected catalogue.

To verify that up to $z\sim3$ our $H$-band-selected catalogue was indeed `complete' down to our stellar mass--completeness limits, we measured stellar mass functions, counting the number of galaxies in bins of redshifts and stellar masses, normalized by the volume probed by the XDF.
Down to our stellar mass--completeness limits, these stellar mass functions agree, within the uncertainties, with the fits inferred by \citet{mortlock_2015} and \citet{davidzon_2017} in the CANDELS and COSMOS fields, respectively.

Finally, we verified that our catalogue did not miss any obvious dust emitters, i.e., galaxies already detected by the ASPECS LP 1.2\,mm survey but not in our $H$-band selected sample.
There are 22 galaxies detected by Gonzalez--Lopez et al.\ (2019b) in the ASPECS LP 1.2\,mm continuum map where the combined PB coverage is better than $75\%$ (i.e., the region of interest of our study; see Section~\ref{subsec:stacking}).
Amongst these sources, 21 have a counterpart in our $H$-band selected catalogue within the synthesized beam of the ASPECS observations (i.e., a radius of $0\farcs6$).
The remaining source is one of the faintest source detected by Gonzalez--Lopez et al.\ (2019b), who also reported no clear NIR counterpart.
Assuming that this source is real and at a redshift of $z\sim0.45$, $z\sim0.80$, $z\sim1.30$, $z\sim1.95$, $z\sim2.75$, $z\sim3.9$ or $z\sim5$, it would increase the cosmic dust mass densities inferred here by $\sim$50\%, 10\%, 3\%, 3\%, 3\%, 10\%, 15\%, respectively, i.e., well within our total uncertainties (see Table~\ref{tab:rho}).
For the cosmic gas mass densities, the impact of this source would depend on its stellar mass: assuming $M_{\ast}=10^{9.5}\,M_\odot$, it implies upwards corrections by $\sim$40\%, 10\%, 6\%, 6\%, 6\%, 10\%, 40\%, respectively;
while assuming $M_{\ast}=10^{10.5}\,M_\odot$, it implies upwards corrections by $\sim$30\%, 7\%, 4\%, 4\%, 4\%, 5\%, 16\%, respectively.
However, because this source is very faint and on the lower end of the SNR ($=4.1$) and `Fidelity' ($=0.78$) selection criteria of Gonzalez--Lopez et al.\ (2019b), it could well be a spurious source (i.e., positive noise peak in the ASPECS LP 1.2\,mm survey).

\section{Method}
\label{sec:methodology}
By combining our stellar mass--complete galaxy sample with our deep ASPECS LP 1.2\,mm survey, we can measure in several redshift bins the total dust and gas mass contained in these galaxies and from that infer their cosmic dust and gas mass densities, knowing the comoving volume probed by our survey. 
In this section, we first summarize the method used to convert observed-frame 1.2\,mm flux densities into dust and gas masses.
We follow by describing the method used to infer the cosmic dust and gas mass densities from these galaxies through stacking.

\subsection{Measuring $M_{\rm dust}$ and $M_{\rm gas}$ from $S_{\rm 1.2\,mm}$}
\label{subsec:measuring_mdust}
In the optically thin approximation, which is almost always valid at the long wavelengths probed by our observations, the dust mass ($M_{\rm dust}$ in $M_{\odot}$) of a galaxy can be inferred using its (sub)millimeter flux density ($S_{\nu_{\rm obs}}$ in Jy) at the observed-frame frequency, $\nu_{\rm obs}=\nu_{\rm rest}/(1+z)$, following, e.g., \citet{kovacs_2010},
\begin{equation}
\label{eq:mdust}
M_{\rm dust} = \frac{5.03\times10^{-31} \cdot S_{\nu_{\rm obs}} \cdot D_{\rm L}^2}{(1+z)^4\cdot B_{\nu_{\rm obs}}(T_{\rm obs}) \cdot \kappa_{\nu_0}}\cdot\bigg(\frac{\nu_0}{\nu_{\rm rest}}\bigg)^{\beta},
\end{equation}
where $B_{\nu_{\rm obs}}(T_{\rm obs})$ is Planck's blackbody function in Jy\,\,sr$^{-1}$ at the observed-frame temperature $T_{\rm obs}$ in Kelvin, which relates to the rest-frame temperature $T$ as $T_{\rm obs}=T/(1+z)$, $D_{\rm L}$ is the luminosity distance in meter, $\beta$ is the dust emissivity spectral index and $\kappa_{\nu_0}$ is the photon cross-section to mass ratio of dust (in m$^2\,$kg$^{-1}$) at rest-frequency $\nu_0$.

As advocated by \citet{scoville_2016}, we used a mass-weighted mean dust temperature of $\langle T\rangle_{\rm M}=T=25\,$K.
This is the mass-dominant dust component of the ISM of galaxies and accounts for the bulk of their Rayleigh-Jeans dust emission \citep{scoville_2016}.
Part of the dust in the ISM can (and will) be at higher temperatures but only in localized regions with a negligible contribution to the global dust mass and Rayleigh-Jeans emission.
A value of 25\,K is supported by \textit{Herschel}-based studies of local and high-redshift galaxies, which find a range of $T\sim15-35\,$K \citep{dunne_2011, dale_2012,auld_2013,magnelli_2014}.
It is further supported at high redshift by the recent work of \citet{kaasinen_2019}, which compared CO-based and dust-based gas measurements at $z\sim2$.
Note that in the Rayleigh-Jeans tail probed by our observations, dust mass estimates vary as $T^{-1}$. 
Thus, a dust temperature range of $T\sim15-35\,$K implies a systematic uncertainty for our dust masses of 25\%--50\%.

As suggested by \citet{scoville_2016}, we used $\beta=1.8$, corresponding to the Galactic measurement made by the \citet{planck_2011a} and is well within the range observed in high-redshift star-forming galaxies \citep[e.g.,][]{chapin_2009a,magnelli_2012}.
Varying $\beta$ within the range predicted by most theoretical models, i.e., $\beta=1.5-2.0$ \citep{draine_2011}, does not impact our results significantly.
Indeed, our dust mass estimates would simply be multiplied by 0.99 (1.00), 1.06 (0.96), 1.14 (0.91), 1.23 (0.87) and 1.32 (0.83), at $z\sim0.45$, $z\sim0.80$, $z\sim1.30$, $z\sim1.95$, and $z\sim2.75$, while using $\beta=1.5\ (2.0)$ instead of $\beta=1.8$, respectively.
Finally, we adopted $\kappa_{\nu_0} = 0.0431\,$m$^2\,$kg$^{-1}$ with $\nu_{0}=352.6\,$GHz \citep[i.e., 850\,$\mu$m;][]{li_2001}.
Interestingly, assuming a typical gas-to-dust mass ratio of 100, this dust mass absorption cross section is within a few percent of the `ISM' mass absorption cross section calibrated by \citet[i.e, their $\alpha_{\rm 850\,\mu m}$]{scoville_2016}.

Finally, following \citet{dacunha_2013}, we corrected our dust mass measurements for the effect of the cosmic microwave background (CMB), which temperature increases as $T_{\rm CMB}(z)=2.73\times(1+z)$.
First, the CMB acts as an additional source of heating of the mass-dominant dust component of galaxies, increasing its temperature from 25\,K at $z=0$ to 25.3\,K at $z=5$ \citep[Equation 12 of][]{dacunha_2013}.
Second, the CMB acts as a background against which we make our measurements, implying an underestimation of the intrinsic flux densities of galaxies \citep[Equation 18 of][]{dacunha_2013}.
Although taken into account, these effects have a relatively minor impact on our results, as they yield upward corrections of our measurements by only 1.01, 1.02, 1.03, 1.04, and 1.07, at $z\sim0.45$, $z\sim0.80$, $z\sim1.30$, $z\sim1.95$, and $z\sim2.75$, respectively.
At $z=3.9$ and $z=5.0$, where our sample cannot be considered as stellar mass complete, these CMB corrections have a value of 1.12 and 1.20, respectively.\\

Dust masses can be converted into gas masses, assuming a gas-to-dust mass ratio, which can be a function of metallicity \citep[e.g.,][]{leroy_2011,eales_2012,magdis_2012,magnelli_2012b,santini_2014,genzel_2015,scoville_2016,tacconi_2018}.
As in \citet{tacconi_2018}, we used a gas-to-dust mass ratio ($\delta_{\rm GDR}$) that correlates nearly linearly with metallicity and is equal to 100 at solar metallicity, $[12+{\rm log(O/H)}]_\odot=8.67$, i.e.,
\begin{equation}
\label{eq:gas_dust_ratio}
\delta_{\rm GDR}=\frac{M_{\rm gas}}{M_{\rm dust}}=10^{(+2-0.85\cdot(12+{\rm log(O/H) _{\rm PP04}}-8.67))},
\end{equation}
where $12+{\rm log(O/H)} _{\rm PP04}$ is the gas phase metallicity adopting the \citet{pettini_2004} scale.
As already mentioned, at solar metallicity (i.e., $\delta_{\rm GDR}=100$), the combination of Equations~\ref{eq:mdust} and \ref{eq:gas_dust_ratio} is equivalent to the method advocated by \citet[]{scoville_2016} for massive galaxies.
In Equation \ref{eq:gas_dust_ratio}, $M_{\rm gas}$ includes both the molecular (H$_{2}$) and atomic (\ion{H}{1}) phases and a standard 36\% mass fraction correction to account for helium.
Equation \ref{eq:gas_dust_ratio} also accounts for the molecular phase that is fully molecular in H$_{2}$ but dissociated in CO \citep[the so-called CO-dark phase; see][]{leroy_2011}.

The metallicity of our galaxies is inferred using the stellar mass-metallicity relation following \citet{tacconi_2018},
\begin{equation}
\label{eq:metallicity}
\begin{array}{l}
    12 + {\rm log(O/H)} _{\rm PP04} = a - 0.087 \times ({\rm log}(M_{\ast})-b)^2,\ {\rm with} \\
    \hspace{0.5cm}a=8.74,\ {\rm and} \\
    \hspace{0.5cm}b = 10.4+4.46\times {\rm log}(1+z) - 1.78\times({\rm log}(1+z))^2.
\end{array}
\end{equation}
We note that while the $\delta_{\rm GDR}$-metallicity relation is believed to be nearly linear at $z\sim0$ down to $12+{\rm log(O/H)}\sim7.9$ \citep{leroy_2011,remy_2014}, at lower metallicities observations suggest that this relation might follow a steeper power-law \citep[$\sim3$;][see also Coogan et al. \citeyear{coogan_2019}]{remy_2014}, yielding larger gas masses per unit dust mass.
In our sample, galaxies with stellar masses close to our stellar mass--completeness limits have metallicities in the range 7.7--7.9.
Consequently, their gas mass could be underestimated by a factor a few when using Equation~\ref{eq:gas_dust_ratio}. 
However, the metallicity at which this transition takes place is very uncertain \citep[$7.94\pm0.47$;][]{remy_2014} and could be irrelevant for our galaxies.
The effect of such steep $\delta_{\rm GDR}$-metallicity relation at very low metallicities on the inferred cosmic gas mass densities is further discussed in Section~\ref{subsec:rho_gas_mass}.

\subsection{Stacking Analysis}
\label{subsec:stacking}
We can measure the total observed--frame 1.2\,mm flux density of a galaxy population knowing their positions within our survey.
To this end, we can start by creating stamps of the ASPECS LP 1.2\,mm survey centered around each of these galaxies.
Then, for a given redshift bin and stellar mass range of interest, we can stack these stamps together, obtaining thereby at the center of this stacked stamp the total emission of a given galaxy population at 1.2\,mm.
Finally, using the comoving volume probed by our survey in this redshift bin, this total emission can be converted into comoving dust and gas mass densities applying the equations given in Section~\ref{subsec:measuring_mdust}.
However, to obtain robust measurements out of this relatively straightforward methodology, precautions must be taken, which are summarized below.

As a first step, to account for the different astrometric solution between the ALMA data and optical/NIR data in the XDF, we applied an astrometry offset ($\Delta$RA$=+0.076\arcsec$ , $\Delta $Dec$=-0.279\arcsec$) to all positions in our master catalogue \citep[][]{rujopakarn_2016,dunlop_2017,cibinel_2017,franco_2018,decarli_2019}.

We then restricted our analysis to the region of the XDF where the combined PB coverage of the ASPECS LP 1.2\,mm survey is better than $75\%$. 
This ensures that our stacking analysis does not include the relatively noisy edges of the survey. 
The surface area of the ASPECS LP 1.2\,mm survey with a combined PB coverage $\,\geq75\%$ is 2.27\,arcmin$^2$.
This is the surface area used to compute the comoving volume probed by the survey in a given redshift bin.

For galaxies that are individually detected (see Section~\ref{sec:data} and Gonzalez--Lopez et al. 2019b) by the ASPECS LP 1.2\,mm survey, stamps were cut out from the `clean' band 6 continuum mosaic. 
For galaxies undetected by the survey, stamps were cut out from the `residual' mosaic, i.e., the `clean' mosaic where detected sources were removed using the `clean' synthesized beam (a 2D Gaussian approximation of the synthesized beam created by the task \textsc{TCLEAN}).
This ensures that we do not count several times the flux density of detected galaxies in the stacking analysis (for stacked sources within 1\arcsec--2\arcsec\ from a detected source) and that the background of our stacked stamps is not dominated by individually detected sources (for stack sources within 2\arcsec\ to 10\arcsec\ from a detected source).
We verified, however, that consistent results are found when stamps for undetected sources were instead cut from the `clean' mosaic.
The size of each stamp is 20\arcsec$\,\times\,$20\arcsec, allowing for an ample number of independent beams in the stacked stamps, crucial for accurate noise estimates.
Note that even at the depth of the ASPECS LP 1.2\,mm survey, confusion noise is negligible (Gonzalez--Lopez et al. 2019b).

The conversion of the observed-frame 1.2\,mm flux density of a galaxy into its dust and gas masses (see Equations~\ref{eq:mdust}, \ref{eq:gas_dust_ratio} and \ref{eq:metallicity}) depends on its redshift and metallicity (and thus stellar mass). 
Therefore, before proceeding with the stacking, we converted each galaxy stamp from Jy\,beam$^{-1}$ to comoving $M_{\rm dust}$ or $M_{\rm gas}$ density units.
To this end, we simply applied the conversion factors to each pixel, normalized by the comoving volume probed by the survey in the redshift bin of interest. 
The final stacked stamps are thus directly in units of $M_\odot$\,Mpc$^{-3}$\,beam$^{-1}$.
\begin{figure*}
\begin{center}
\includegraphics[width=\linewidth]{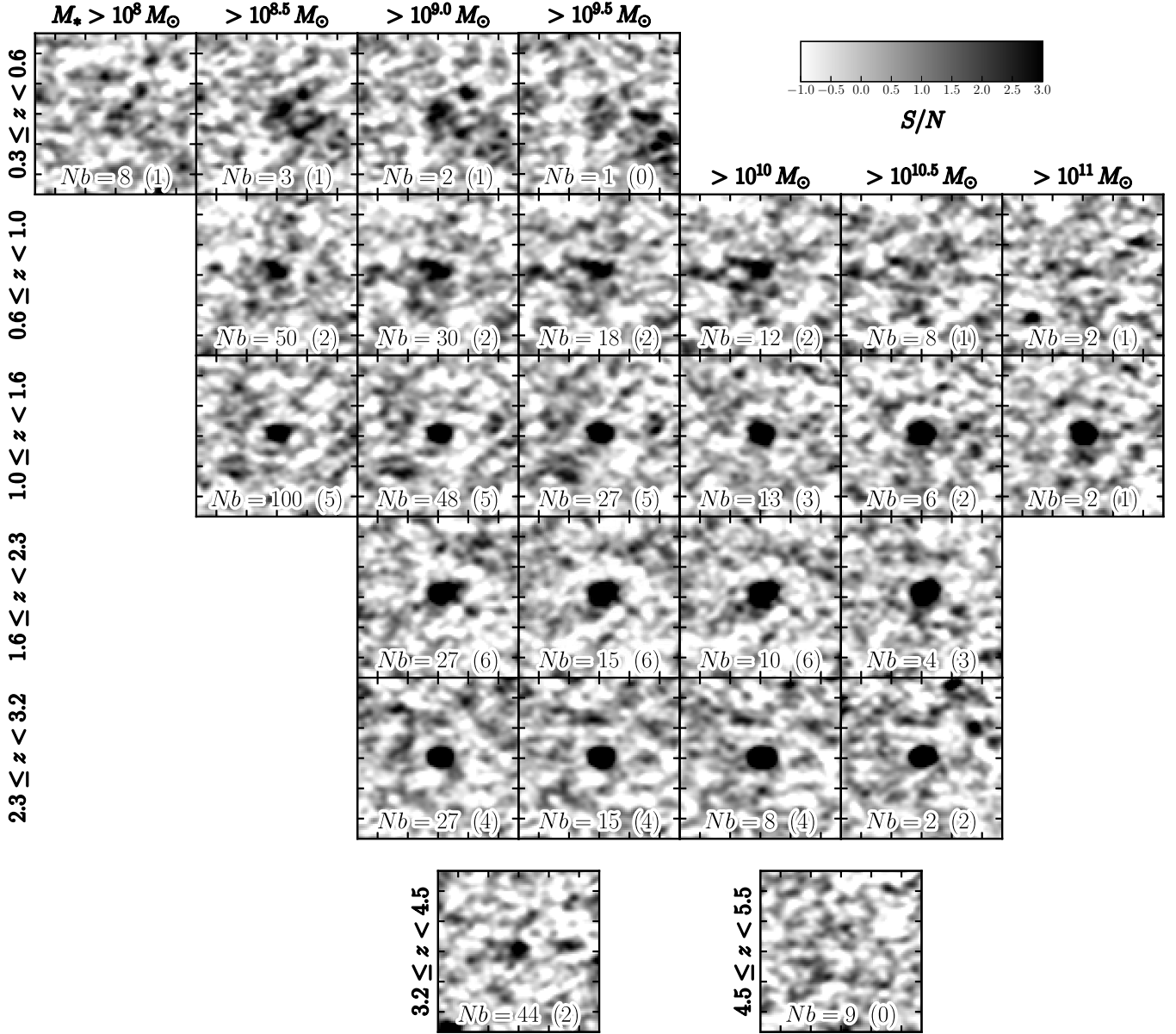}
\caption{\label{fig:stamps}
        $16\arcsec\times16\arcsec$ zoom-in on the cumulative stacked stamps corresponding to the comoving dust mass density in galaxies in different redshift bins and above a given stellar mass, i.e., $\rho_{\rm dust}(M_\ast>M,z)$.
        Because these are cumulative stacked stamps, individual panels are not independent.
        To ease the assessment of the detection significance, the color-scale of each stamp is set to vary as $S_{\rm pix}/N_{\rm pix}$, where $S_{\rm pix}$ is the pixel signal and $N_{\rm pix}$ is the standard deviation of the pixel distribution (both in units of $M_\odot$\,Mpc$^{-3}$\,beam$^{-1}$).
        The number of source stacked ($Nb$) is indicated in each stamp, while in parenthesis is the number of sources among them which are individually detected at 1.2\,mm.
        At $z<3.2$, we only show stellar masses not affected by incompleteness (see Figure~\ref{fig:lh_vs_mass}).
        At $z>3.2$, where our $H$-band-selected sample cannot be considered as stellar mass-selected, we stacked all $z>3.2$ galaxies together, irrespective of their stellar masses.
		}
\end{center}
\vspace{0.2cm}
\end{figure*}
\begin{figure*}
\begin{center}
\includegraphics[width=\linewidth]{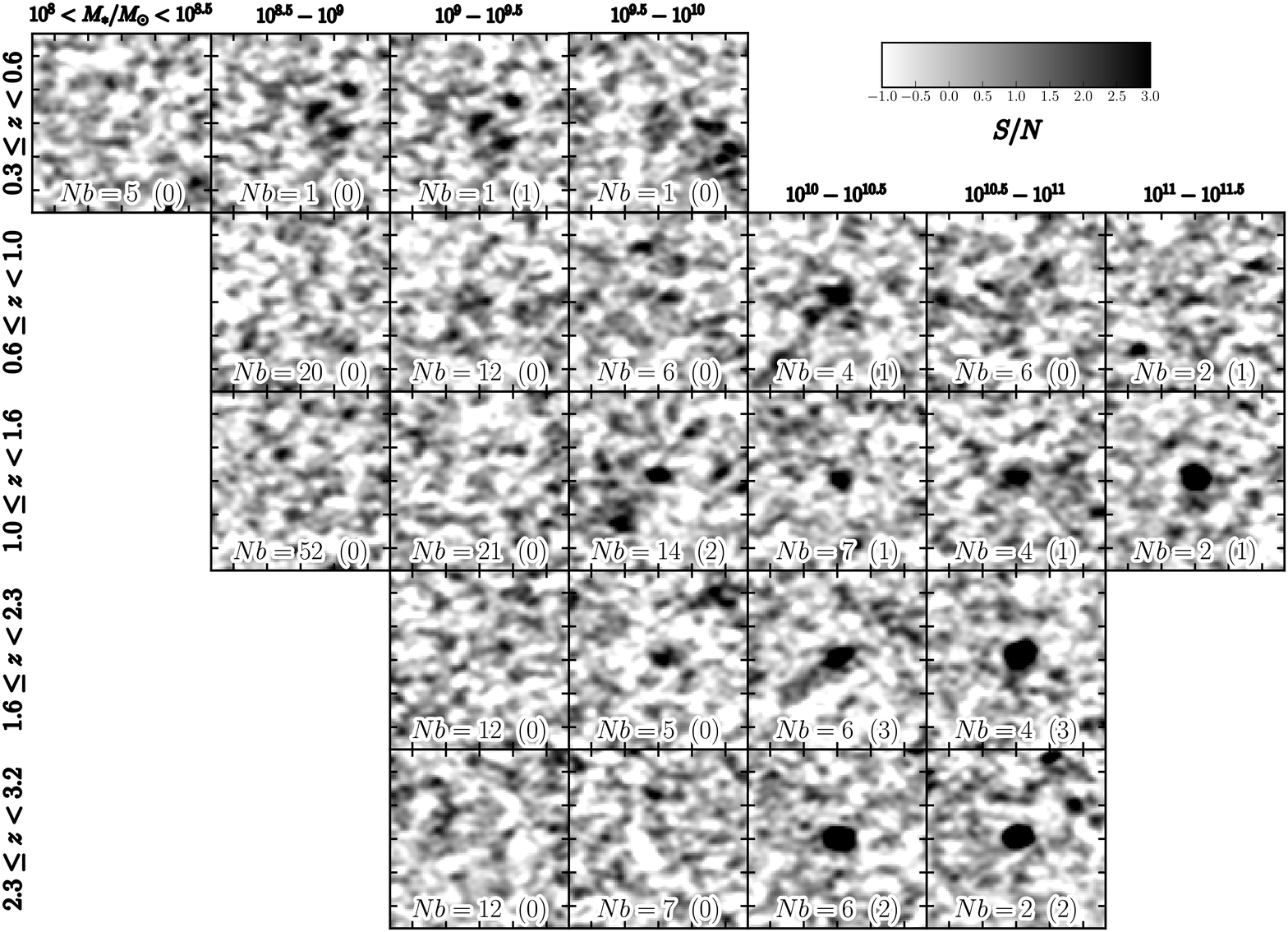}
\caption{\label{fig:stamps_diff}
        Same as Figure~\ref{fig:stamps} but for the differential stacked stamps, i.e., the cosmic dust mass density in galaxies in a given redshift and stellar mass bin, $\rho_{\rm dust}(M_{\ast} \in [M\pm0.25 {\rm dex}],z)$.
        In the first row, the second and third panels look very similar.
        It corresponds to a close pair of galaxies, both at $z=0.458$ and separated by only $1\farcs7$, and falling into different stellar mass bins. 
		}
\end{center}
\vspace{0.2cm}
\end{figure*}

The signal in the stacked stamps was measured as follows. 
First, we measured at their center the signal within an aperture with a 0.6$\times$FWHM$^{\rm beam}$ radius, optimized for point-source detection (i.e., equivalent to a matched--filter).
This signal ($S_{\rm aper}$) was then compared to the noise ($N_{\rm aper}$), defined as the standard deviation of the signal distribution within 200 apertures randomly positioned in the stacked stamp.
If $S_{\rm aper}/N_{\rm aper}\geq5$, we fitted the stacked stamp with a 2D Gaussian function.
The amplitude ($A_{\rm fit}$) as well as the minor ($\theta_{\rm min}^{\rm fit}$) and major ($\theta_{\rm maj}^{\rm fit}$) FWHM of this 2D Gaussian function centered on the stacked stamp were left as free parameters, accommodating thereby any possible astrometric mismatch between the near-infrared and millimeter centers of these galaxies.
The comoving mass density (i.e., $\rho_{\rm dust}$ or $\rho_{\rm gas}$, depending on which conversion was applied beforehand to the 1.2\,mm stacked stamps) and its associated measurement uncertainty ($\sigma_\rho^{\rm m}$) were then derived using the standard formulae, 
\begin{equation}
\label{eq:flux}
    \rho_{\rm dust}\ {\rm or}\ \rho_{\rm gas}\  [M_{\odot}{\rm\,Mpc^{-3}]} = A_{\rm fit}\cdot\frac{\theta_{\rm min}^{\rm fit} \cdot \theta_{\rm maj}^{\rm fit}}{\theta_{\rm min}^{\rm beam} \cdot \theta_{\rm maj}^{\rm beam}},
\end{equation}
\begin{equation}
\label{eq:error}
    \sigma_\rho^{\rm m}\  [M_{\odot}{\rm\,Mpc^{-3}]} = \sigma_{\rm pixel}\cdot\sqrt{\frac{\theta_{\rm min}^{\rm fit} \cdot \theta_{\rm maj}^{\rm fit}}{\theta_{\rm min}^{\rm beam} \cdot \theta_{\rm maj}^{\rm beam}}},
\end{equation}
where $\theta_{\rm min}^{\rm beam}$ and $\theta_{\rm maj}^{\rm beam}$ are the minor and major FWHM of the synthesized beam; and $\sigma_{\rm pixel}$ is the standard deviation of the stacked stamp pixel distribution. 

In stacked stamps where $S_{\rm aper}/N_{\rm aper}<5$, the comoving mass density ($\rho_{\rm dust}$ or $\rho_{\rm gas}$) was instead given by the total signal in an aperture with a 1.0$\times$FWHM$^{\rm beam}$ radius, divided by the area of the synthesized beam in units of pixel; and $\sigma_\rho^{\rm m}$ was given by the standard deviation of the signal distribution within 200 apertures randomly positioned in the stacked stamp, divided by the area of the synthesized beam in units of pixel.
This particular radius was chosen to be large enough to encompass the total signal in the stacked stamp and to provide consistent signal measurements with respect to our 2D Gaussian fits when applied on clear detections (i.e., when $S_{\rm aper}/N_{\rm aper}>5$).

The total uncertainty ($\sigma_{\rho}^{\rm tot}$) was defined as the quadratic sum of the measurement uncertainty ($\sigma_\rho^{\rm m}$) and the Poissonian uncertainty ($\sigma_\rho^{\rm Poisson}$).
The former accounts for the noise in the stacked stamp (i.e., detection significance), while the latter accounts for the uncertainty due to the low number of galaxies stacked in each bin (i.e., $\propto\sqrt{1/Nb}$).

We performed the stacking analysis in five redshift bins in which our galaxies can be considered as stellar mass-selected, i.e., $0.3\leq z<0.6$, $0.6\leq z<1.0$, $1.0\leq z<1.6$, $1.6\leq z<2.3$, and $2.3\leq z<3.2$ (Section~\ref{sec:sample} and Figure~\ref{fig:lh_vs_mass}).
These redshift bins were chosen to sample the cosmic history with a reasonably large number of sources ($>8$) per $<\,2$ Gyr look--back time intervals.
In these redshift bins, we stacked all galaxies with a stellar mass greater than a given threshold, starting at $10^{11}\,M_{\odot}$ and decreasing it in steps of 0.5\,dex down to our stellar mass-completeness limit.

The stacked stamps obtained for the dust are shown in Figure~\ref{fig:stamps}, while the measured cosmic dust and gas mass densities, i.e., $\rho_{\rm dust}(M_\ast>M,z)$ and $\rho_{\rm gas}(M_\ast>M,z)$, and associated uncertainties are given in Table~\ref{tab:rho} and \ref{tab:rho_gas}.

From right to left in Figure~\ref{fig:stamps}, the stack stamps include ever less massive galaxies.
Because these are cumulative stacked stamps, individual panels are not independent.
At $z>0.6$, the stacking analysis yields significant detections ($\gtrsim5\sigma_\rho^{\rm m}$) for most stellar mass thresholds down to our stellar mass--completeness limits.
At $0.3\leq z<0.6$, the stacking analysis yields mostly tentative detections ($<3\sigma_\rho^{\rm m}$) but these measurements provide meaningful constraints on the cosmic dust mass density at this redshift and above these stellar mass thresholds (see Section~\ref{subsec:rho_dust_mass}).
Note that the counter-intuitive increase in detection significance from $0.3\leq z<0.6$ (top row in Figure~\ref{fig:stamps}) to $2.3\leq z<3.2$ (bottom row) is due to (i) the intrinsically lower dust mass content in low redshift galaxies (see Section~\ref{subsec:rho_dust_mass}); (ii) the smaller cosmic volume and thus the fewer number of galaxies in the lower redshift bins (see Figure~\ref{fig:lh_vs_mass}); and (iii) to the fact that on the Rayleigh-Jeans tail and at a given observed wavelength, low-redshift galaxies are not significantly brighter per unit dust mass than high-redshift galaxies, because of the so-called negative $K$-correction \citep[see, e.g., Figure 2 in ][]{scoville_2016}.
\begin{table*}
\begin{center}
\begin{rotatetable*}
\caption{\label{tab:rho} Cosmic dust mass density in galaxies in different redshift bins and above a given stellar mass, i.e., $\rho_{\rm dust}(M_\ast>M,z)$. 
The total uncertainties correspond to the quadratic sum of the measurement uncertainties ($\sigma_\rho^{\rm m}$) and the Poissonian uncertainties ($\sigma_\rho^{\rm Poisson}$).
The measurement uncertainties are provided in parentheses and should be used to assess the detection significance in the stacked stamps.
$Nb$ gives the number of stacked galaxies.
In squared brackets, we provide the cosmic dust mass densities inferred by fitting the variation of $\rho_{\rm dust}(M_\ast>M,z)$ using the SMF of \citet{mortlock_2015} and solving for $f_{\rm dust}$ assuming $f_{\rm dust}=M_{\rm dust} / M_{\ast}=A\times (M\,/\,10^{10.7}\,M_\odot)^{B}$ (see Section~\ref{subsec:fdust}).
At $3.2\leq z<4.5$ and $4.5\leq z<5.5$, where the sample cannot be considered as stellar mass--selected, we summed the contribution of all galaxies with $H\leq27$, irrespective of their stellar masses.
Differential measurements can be inferred by subtracting accordingly the values of interest.}
\begin{tabular}{ c c c c c c c c } 
\hline \hline
\rule{0pt}{3ex} & \multicolumn{7}{c}{$\rho_{\rm dust}(M_\ast>M,z)$}\\
 Redshift & \multicolumn{7}{c}{[$\times 10^5$ $M_\odot$\,Mpc$^{-3}$]}\\
\cline{2-8}
 \rule{0pt}{3ex}bin & $>10^{8}\,M_\odot$ & $>10^{8.5}\,M_\odot$ & $>10^{9}\,M_\odot$ & $>10^{9.5}\,M_\odot$ & $>10^{10}\,M_\odot$ & $>10^{10.5}\,M_\odot$ & $>10^{11}\,M_\odot$ \\
\hline
\rule{0pt}{5ex} & $Nb=8$   &   $Nb=3$ &      $Nb=2$ & $Nb=1$ \\
\rule{0pt}{3ex}$0.3\leq z<0.6$ & $1.1\pm0.8$ ($0.5$)   &   $1.0\pm0.7$ ($0.3$) &      $0.8\pm0.6$  ($0.2$) & $0.3\pm0.3$ ($0.2$) & $\cdots$ & $\cdots$ & $\cdots$ \\
\rule{0pt}{3ex} & $[1.2^{+0.6}_{-0.5}]$ & & $[0.5^{+0.3}_{-0.2}]$ & & $[0.2^{+0.3}_{-0.1}]$\\
\cline{2-8}
\rule{0pt}{5ex} &    &   $Nb=50$ &  $Nb=30$ & $Nb=18$ & $Nb=12$ & $Nb=8$ & $Nb=2$ \\
\rule{0pt}{3ex}$0.6\leq z<1.0$ & $\cdots$   &   $3.0\pm0.9$ ($0.7$) &      $2.7\pm0.8$ ($0.6$) & $1.9\pm0.7$ ($0.4$)  & $1.4\pm0.6$ ($0.3$)  & $0.6\pm0.3$ ($0.3$)  & $0.2\pm0.2$ ($0.1$) \\
\rule{0pt}{3ex} & $[3.3^{+1.6}_{-1.2}]$ & & $[2.6^{+0.7}_{-0.6}]$ & & $[1.4^{+0.4}_{-0.6}]$\\
\cline{2-8}
\rule{0pt}{5ex} &    &   $Nb=100$ &  $Nb=48$ & $Nb=27$ & $Nb=13$ & $Nb=6$ & $Nb=2$ \\
\rule{0pt}{3ex}$1.0\leq z<1.6$ & $\cdots$   &   $4.7\pm1.4$ ($0.5$) &      $4.3\pm1.3$ ($0.3$) & $4.2\pm1.3$ ($0.3$)  & $3.1\pm1.3$ ($0.2$)  & $2.4\pm1.2$ ($0.1$)  & $1.6\pm1.2$ ($0.07$) \\
\rule{0pt}{3ex} & $[4.3^{+1.8}_{-1.4}]$ & & $[4.2^{+1.1}_{-1.3}]$ & & $[3.6^{+1.1}_{-1.3}]$\\
\cline{2-8}
\rule{0pt}{5ex} &    &    &  $Nb=27$ & $Nb=15$ & $Nb=10$ & $Nb=4$  \\
\rule{0pt}{3ex}$1.6\leq z<2.3$ & $\cdots$   &   $\cdots$ &      $3.5\pm1.1$  ($0.2$) & $3.3\pm1.1$ ($0.1$)  & $3.0\pm1.1$ ($0.1$)  & $2.1\pm1.0$ ($0.07$)  & $\cdots$ \\
\rule{0pt}{3ex} & $[4.0^{+1.5}_{-1.3}]$ & & $[3.6^{+0.9}_{-1.0}]$ & & $[2.6^{+0.9}_{-0.1}]$\\
\cline{2-8}
\rule{0pt}{5ex} &    &    &  $Nb=27$ & $Nb=15$ & $Nb=8$ & $Nb=2$  \\
\rule{0pt}{3ex}$2.3\leq z<3.2$ & $\cdots$   &   $\cdots$ &      $3.1\pm1.1$  ($0.2$) & $3.0\pm1.1$ ($0.1$)  & $2.9\pm1.1$ ($0.1$)  & $1.1\pm0.8$ ($0.04$)  & $\cdots$ \\
\rule{0pt}{3ex} & $[3.8^{+2.3}_{-1.2}]$ & & $[3.2^{+0.9}_{-0.7}]$ & & $[2.1^{+0.6}_{-0.4}]$\\
\cline{2-8}

\rule{0pt}{5ex} $3.2\leq z<4.5$ &  \multicolumn{7}{c}{$Nb=44$\ \ \ \ \ \ \ $0.6\pm0.2$ ($0.1$)} \\

\rule{0pt}{5ex} $4.5\leq z<5.5$ &  \multicolumn{7}{c}{$Nb=9$\ \ \ \ \ \ \ $0.05\pm0.07$ ($0.07$)} \\
\hline
\hline
\end{tabular}
\end{rotatetable*}
\end{center}
\end{table*}
\begin{table*}
\begin{rotatetable*}
\begin{center}
\caption{\label{tab:rho_gas} Same as Table~\ref{tab:rho} but for the cosmic gas mass density in galaxies}
\begin{tabular}{ c c c c c c c c } 
\hline \hline
\rule{0pt}{3ex} & \multicolumn{7}{c}{$\rho_{\rm gas}(M_\ast>M,z)$}\\
 Redshift & \multicolumn{7}{c}{[$\times 10^7$ $M_\odot$\,Mpc$^{-3}$]}\\
\cline{2-8}
 \rule{0pt}{3ex}bin & $>10^{8}\,M_\odot$ & $>10^{8.5}\,M_\odot$ & $>10^{9}\,M_\odot$ & $>10^{9.5}\,M_\odot$ & $>10^{10}\,M_\odot$ & $>10^{10.5}\,M_\odot$ & $>10^{11}\,M_\odot$ \\
\hline
\rule{0pt}{5ex} & $Nb=8$   &   $Nb=3$ &      $Nb=2$ & $Nb=1$ \\
\rule{0pt}{3ex}$0.3\leq z<0.6$ & $1.8\pm1.6$ ($1.3$)   &   $1.4\pm1.0$ ($0.4$) &      $1.0\pm0.8$  ($0.3$) & $0.3\pm0.3$ ($0.2$) & $\cdots$ & $\cdots$ & $\cdots$ \\
\rule{0pt}{3ex} & $[2.0^{+1.1}_{-1.1}]$ & & $[0.8^{+0.3}_{-0.4}]$ & & $[0.2^{+0.2}_{-0.2}]$\\
\cline{2-8}
\rule{0pt}{5ex} &    &   $Nb=50$ &  $Nb=30$ & $Nb=18$ & $Nb=12$ & $Nb=8$ & $Nb=2$ \\
\rule{0pt}{3ex}$0.6\leq z<1.0$ & $\cdots$   &   $3.8\pm1.7$ ($1.5$) &     $3.3\pm1.2$  ($1.0$) & $2.2\pm0.8$ ($0.5$)  & $1.4\pm0.6$ ($0.3$)  & $0.6\pm0.3$ ($0.2$)  & $0.2\pm0.2$ ($0.1$) \\
\rule{0pt}{3ex} & $[5.0^{+2.1}_{-2.6}]$ & & $[3.3^{+0.8}_{-1.3}]$ & & $[1.4^{+0.5}_{-0.5}]$\\
\cline{2-8}
\rule{0pt}{5ex} &    &   $Nb=100$ &  $Nb=48$ & $Nb=27$ & $Nb=13$ & $Nb=6$ & $Nb=2$ \\
\rule{0pt}{3ex}$1.0\leq z<1.6$ & $\cdots$   &   $6.4\pm2.0$ ($1.5$) &      $5.3\pm1.5$  ($0.6$) & $5.1\pm1.4$ ($0.4$)  & $3.3\pm1.2$ ($0.2$)  & $2.3\pm1.1$ ($0.1$)  & $1.5\pm1.1$ ($0.07$) \\
\rule{0pt}{3ex} & $[5.6^{+3.3}_{-1.7}]$ & & $[5.4^{+1.7}_{-1.4}]$ & & $[4.1^{+0.9}_{-1.7}]$\\
\cline{2-8}
\rule{0pt}{5ex} &    &    &  $Nb=27$ & $Nb=15$ & $Nb=10$ & $Nb=4$  \\
\rule{0pt}{3ex}$1.6\leq z<2.3$ & $\cdots$   &   $\cdots$ &      $5.2\pm1.5$  ($0.5$) & $4.5\pm1.4$ ($0.3$)  & $3.9\pm1.3$ ($0.2$)  & $2.4\pm1.2$ ($0.1$)  & $\cdots$ \\
\rule{0pt}{3ex} & $[5.6^{+1.8}_{-1.2}]$ & & $[5.0^{+0.9}_{-0.7}]$ & & $[3.6^{+0.8}_{-0.6}]$\\
\cline{2-8}
\rule{0pt}{5ex} &    &    &  $Nb=27$ & $Nb=15$ & $Nb=8$ & $Nb=2$  \\
\rule{0pt}{3ex}$2.3\leq z<3.2$ & $\cdots$   &   $\cdots$ &      $5.6\pm1.9$  ($0.5$) & $5.3\pm1.8$ ($0.2$)  & $5.1\pm1.8$ ($0.1$)  & $1.4\pm1.0$ ($0.04$)  & $\cdots$ \\
\rule{0pt}{3ex} & $[9.2^{+2.9}_{-3.9}]$ & & $[6.6^{+0.1}_{-1.8}]$ & & $[3.5^{+0.8}_{-0.7}]$\\
\cline{2-8}

\rule{0pt}{5ex} $3.2\leq z<4.5$ &  \multicolumn{7}{c}{$Nb=44$\ \ \ \ \ \ \ $2.4\pm1.6$ ($1.3$)} \\

\rule{0pt}{5ex} $4.5\leq z<5.5$ &  \multicolumn{7}{c}{$Nb=9$\ \ \ \ \ \ \ $1.3\pm1.8$ ($1.7$)} \\
\hline
\hline
\end{tabular}
%\end{center}
%\textbf{Notes.}\\
%$^{\rm{(a)}}$ \\
%$^{\rm{(b)}}$
\end{center}
\end{rotatetable*}
\end{table*}

At $z>3.2$, where the sample cannot be considered as stellar mass--complete, we divided it into two redshift bins, $3.2\leq z<4.5$ and $4.5\leq z<5.5$, and stacked all galaxies with $H\leq27$ (see Figure~\ref{fig:stamps}).
Due to stellar mass-incompleteness, these measurements only provide upper limits on the cosmic dust and gas mass densities in galaxies at these redshifts.
However, because our 1.2\,mm-to-gas mass conversion requires accurate metallicity and consequently stellar mass measurements, the inferred upper limits on the cosmic gas mass density at $z>3.2$ are affected by systematics not included in our total uncertainties.\\

Figure~\ref{fig:stamps_diff} shows the differential stacked stamps for the dust, i.e., the cosmic dust mass density in galaxies in a given redshift and stellar mass bins, $\rho_{\rm dust}(M_{\ast} \in [M\pm0.25 {\rm dex}],z)$.
In most redshift bins, we obtain significant detections for our massive stellar mass bins while low mass galaxies are mostly undetected.
This implies that, at our stellar mass-completeness limits, the significant detections observed in our cumulative stacked stamps are dominated by the contribution of massive galaxies ($\gtrsim10^{9.5-10}\,M_\odot$).
The fact that $\lesssim10^{9.5}\,M_\odot$ galaxies only mildly contribute to the cosmic dust (and gas) mass density in galaxies is the main result of the paper \citep[see also][]{dunlop_2017} and is further illustrated and discussed in Sections~\ref{sec:rho_dust} and \ref{sec:rho_gas}.
We note that in fact at $0.3\leq z<0.6$, $0.6\leq z<1.0$, $1.0\leq z<1.6$, $1.6\leq z<2.3$, and $2.3\leq z<3.2$, the sources individually detected in the ASPECS LP 1.2\,mm map contribute for about 30\%, 20\%, 70\%, 80\%, and 80\% of the cumulative measurements at our stellar mass-completeness limits, respectively.\\

Finally, to check the robustness of these results, we repeated these measurements using a bootstrap analysis.
For this, we made 200 realizations of the stacking analysis, using for each realization a different sample, drawn from the original one, with the same number of sources, but allowing for replacement (i.e., a galaxy can be picked several times).
We found that the mean values and standard deviations over these 200 realizations were fully consistent with the values and total uncertainties quoted in Tables~\ref{tab:rho} and \ref{tab:rho_gas}.
Using the same approach, we also test our method against positional offset between the $H$-band and 1.2\,mm emission.
For that, in each bootstrap iteration, we randomly draw the position of each sources within 2D Gaussian functions centered on their original position and with a standard deviation of $0\farcs6$ \citep[typical $H$-band to ALMA offset for $z\sim2$ star-forming galaxies;][]{elbaz_2018}.
Again, the mean values and standard deviations inferred over these 200 realizations were fully consistent with those quoted in Tables~\ref{tab:rho} and \ref{tab:rho_gas}.
As a last sanity check, we randomized the position of the galaxies in the sample and repeated the stacking analysis.
We only obtained non-detections.

\begin{figure*}
\begin{center}
\includegraphics[width=\linewidth]{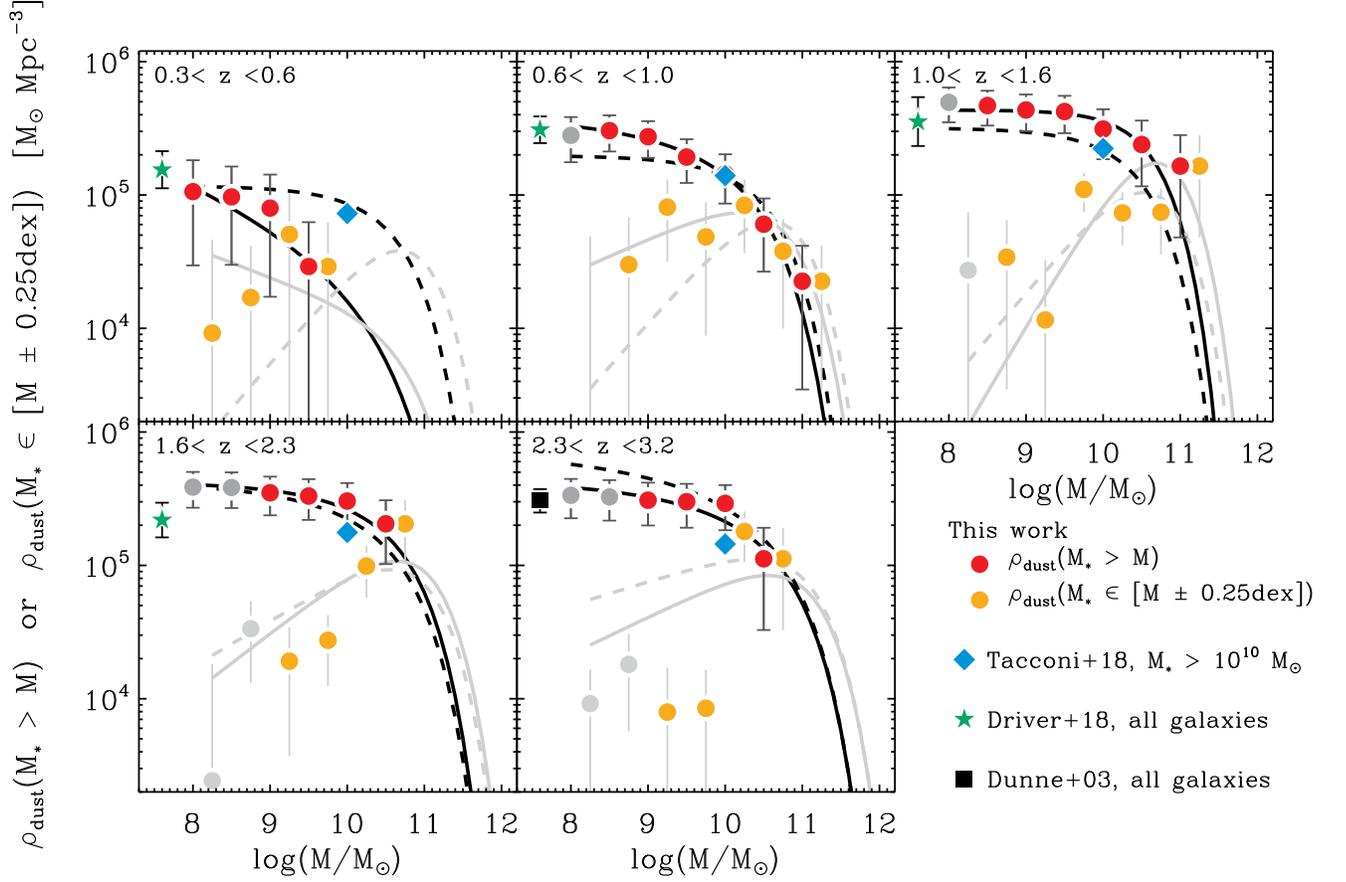}	
\caption{\label{fig:Mdust_vs_Mstar}
        Comoving dust mass density in galaxies for different redshift bins.
        Red and dark gray circles correspond to the cosmic dust mass densities in galaxies with stellar masses $>M$ (i.e., $\rho_{\rm dust}(M_\ast>M,z)$; cumulative stacking), above and below our stellar mass-completeness limits (i.e., $M_{\rm limit}$; see Section~\ref{sec:sample}), respectively.
        Yellow and light gray circles show the cosmic dust mass densities in galaxies with stellar masses $\in [M\pm0.25 {\rm dex}]$ (i.e., $\rho_{\rm dust}(M_{\ast} \in [M\pm0.25 {\rm dex}],z)$; differential stacking), above and below our stellar mass-completeness limits, respectively.
        Some of these differential datapoints have SNR$\,<3$ (see Figure~\ref{fig:stamps_diff}).
        The black solid lines show the best-fits of $\rho_{\rm dust}(M_\ast>M,z)$ using the SMF of \citet{mortlock_2015} and solving for $f_{\rm dust}$ assuming $f_{\rm dust}=M_{\rm dust} / M_{\ast}=A\times (M\,/\,10^{10.7}\,M_\odot)^{B}$ (see Section~\ref{subsec:fdust}). 
        The black dashed lines show the best-fits of $\rho_{\rm dust}(M_\ast>M,z)$ when all redshift bins are fitted simultaneously solving for $f_{\rm dust}$ assuming ${\rm log}(f_{\rm dust})=C+D\times{\rm log}(1+z) + B\times({\rm log}(M_{\ast}/M_\odot)-10.7)$.
        The light gray solid and dashed lines show the exact same best-fits but displayed in differential form, i.e., within stellar mass bins which are 0.5 dex wide.
        Blue diamonds show predictions for galaxies with $>\,$10$^{10}\,M_\odot$ using $f_{\rm gas}$($M,z$) for main-sequence galaxies from \citet{tacconi_2018}, the SMF of \citet{mortlock_2015}, and assuming a gas-to-dust ratio of 100.
        Green stars present the \textit{total} comoving dust mass density in galaxies measured by \citet{driver_2018}, applying the energy-balance code \texttt{MAGPHYS} to hundreds of thousands of galaxies in the GAMA/G10-COSMOS/3D-\textit{HST} surveys.
        The black square show the \textit{total} comoving dust mass density of galaxies measured by \citet{dunne_2003} using single-dish (sub)millimeter-selected galaxies.
        }
\end{center}
\vspace{0.2cm}
\end{figure*}

\section{The cosmic dust mass density in galaxies}
\label{sec:rho_dust}
\subsection{$\rho_{\rm dust}(M_\ast>M,z)$ vs. $M_\ast$}
\label{subsec:rho_dust_mass}
The evolution of the comoving dust mass density in galaxies with stellar masses $>M$, i.e., $\rho_{\rm dust}(M_\ast>M,z)$, as derived from the stacking above, is shown in Figure~\ref{fig:Mdust_vs_Mstar} and tabulated in Table~\ref{tab:rho}.
At $z>0.6$, $\rho_{\rm dust}(M_\ast>M,z)$ grows rapidly as $M$ decreases down to $\sim10^{10}\,M_{\odot}$, but this growth significantly slows down as $M$ decreases to even lower stellar masses.
At $z<0.6$, the measurements are unfortunately too uncertain to fully confirm the existence of such trend.
The flattening of $\rho_{\rm dust}(M_\ast>M,z)$ at low $M$, which happens well above our stellar mass-completeness limits, implies that (i) at our stellar mass-completeness limits our analysis already accounts for most of the dust in the universe locked in galaxies; and (ii) the contribution of low mass galaxies ($\lesssim10^{9.5-10}\,M_{\odot}$) to the total cosmic dust mass density in galaxies becomes rapidly negligible.
This latter finding is clearly illustrated by the differential measurements (i.e., $\rho_{\rm dust}(M_{\ast} \in [M\pm0.25 {\rm dex}],z)$), which peaks around $10^{10-10.5}\,M_{\odot}$ in most redshift bins.
This characteristic stellar mass of $10^{10-10.5}\,M_{\odot}$ where most of the dust in galaxies is locked, is consistent with the characteristic stellar mass of star forming galaxies where most of new stars were formed out to $z\sim3$ \citep[$10^{10.6\pm0.4}\,M_\odot$;][]{karim_2011}.
From a more observational point of view, we note that our results are also consistent with the flattening of the cumulative 1.2\,mm number counts found by Gonzalez--Lopez et al.\ (2019b) at the unparalleled depth of the ASPECS LP 1.2\,mm survey.

In the sample, about 10\%, 20\%, 5\%, 5\%, and 6\% of the galaxies at $0.3\leq z<0.6$, $0.6\leq z<1.0$, $1.0\leq z<1.6$, $1.6\leq z<2.3$, and $2.3\leq z<3.2$, are classified as quiescent using a standard $UVJ$ selection method \citep[e.g.,][]{mortlock_2015}, respectively.
Excluding these galaxies from the stacking analysis does not significantly change our $\rho_{\rm dust}(M_\ast>M,z)$ estimates.
At our stellar mass-completeness limits, $\rho_{\rm dust}(M_\ast>M,z)$ decreases by $<5\,$\% in the first three redshift bins and by $\sim13\%$ in the highest two.
Considering that part of this decrease can actually be due to dusty star-forming galaxies contaminating the quiescent $UVJ$ selection \citep{mortlock_2015,schreiber_2015}, we conclude that the bulk of the dust in galaxies resides in star-forming galaxies. 

As ours is the first study, to our knowledge, to constrain the evolution of $\rho_{\rm dust}(M_\ast>M,z)$ with stellar mass, we can thus only compare our results to the redshift evolution of the \textit{total} $\rho_{\rm dust}$ in galaxies\footnote{as opposed to literature measurements that could include a significant contribution from the dust in the circumgalactic medium, e.g., \citet{debernardis_2012,menard_2012,thacker_2013}.} derived by \citet{dunne_2003,dunne_2011} and \citet{driver_2018}.
For clarity, in Figure~\ref{fig:Mdust_vs_Mstar} we displayed those $\rho_{\rm dust}$ measurements at a stellar mass of $10^{7.6}\,M_\odot$.
\citet{driver_2018} measured the $\rho_{\rm dust}$ by fitting the optical-to-far-infrared photometry of 200,000 galaxies using the energy-balance code \texttt{MAGPHYS}.
Although these estimates relied mostly on dust masses extrapolated from optical dust extinction due to the relatively limited depth of the \textit{Herschel} observations used by Driver et al., they are in very good agreement with our measurements.
%The consistency between Driver et al.'s optically-based and our millimeter-based dust mass estimates suggests that the bulk of the dust resides in regions of galaxies which are not heavily obscured at optical wavelengths and thus reasonably well modelled by the energy-balance code \texttt{MAGPHYS}. 
%This is consistent with the fact that extreme starbursts account for only $\sim5\%$ by number of the star-forming galaxy population in the redshift range probed by our study \citep{rodighiero_2011,schreiber_2015}.
Similarly, we find good agreement with \citet{dunne_2003,dunne_2011}, who measured $\rho_{\rm dust}$ by integrating dust mass functions constrained from ground-based single-dish submillimeter observations and assuming $T=25\,$K and $\beta=2$.
This agreement is somewhat surprising, considering that the faint-end slopes of these dust mass functions at high-redshifts were only loosely constrained by those observations and thus fixed to that observed at $z<0.1$.

The flattening of $\rho_{\rm dust}(M_\ast>M,z)$ toward low stellar masses, together with the agreements with the \textit{total} $\rho_{\rm dust}$ found by \citet{dunne_2003,dunne_2011} and \citet{driver_2018}, suggest that at the stellar mass-completeness limits of our study, we already have accounted for most of the dust in the universe locked in galaxies.

\subsection{The dust-to-stellar mass ratio of star-forming galaxies}
\label{subsec:fdust}
At a given redshift, the variations of $\rho_{\rm dust}(M_\ast>M,z)$ with $M$ can be modelled using the stellar mass function (i.e., SMF$(M,z)$) and the mean dust-to-stellar mass ratio of galaxies (i.e, $f_{\rm dust}(M,z)$),
\begin{equation}
    \label{eq:rho_dust_model}
    \rho_{\rm dust}(M_\ast>M,z)=\int_{M}^{\infty} f_{\rm dust}(M,z)\times {\rm SMF}(M,z)\ dM.
\end{equation}
Because the SMF of galaxies is well known up to $z\sim3$ \citep[e.g.,][]{mortlock_2015}, one can solve for $f_{\rm dust}(M,z)$ by fitting $\rho_{\rm dust}(M_\ast>M,z)$.
We excluded from the fits the measurements below our stellar mass-completeness limits and used the SMF of star-forming galaxies inferred by \citet{mortlock_2015}\footnote{the SMF inferred from our $H$-band-selected sample is consistent with that inferred by \citet{mortlock_2015}. However, this $H$-band-selected sample is too small to robustly re-derived the SMF up to $z\sim3$. Thus, we used instead that from \citet{mortlock_2015}.}, i.e., 
\begin{equation}
    {\rm SMF}(M,z)=\phi^* \cdot {\rm ln}(10) \cdot \bigg(\frac{M}{M^*}\bigg)^{1+\alpha} \cdot e^{-M/M^*}, 
\end{equation}
where $\phi^*$ is the normalization of the Schechter function, $M^*$ is its turnover mass, and $\alpha$ is its low-mass end slope (Table~\ref{tab:schetcher}).
We did not use the SMF that includes quiescent galaxies, as the contribution of those to $\rho_{\rm dust}(M_\ast>M, z)$ has been shown to be negligible (see Section~\ref{subsec:rho_dust_mass}).
$f_{\rm dust}(M,z)$ is thus the mean stellar-to-dust mass ratio of star-forming galaxies. 
\begin{table}[]
    \centering
        \caption{The single Schechter parameters for the star-forming galaxy SMFs, as found in \citet[]{mortlock_2015}.}
    \begin{tabular}{c c c c}
        \hline
        \hline
        Redshift bin & log\,$M^*$ & log\,$\phi^*$ & $\alpha$ \\
        \hline
         $0.3<z<0.5$ & 10.83 & -3.31 & -1.41 \\
         $0.5<z<1.0$ & 10.77 & -3.28 & -1.45 \\ 
         $1.0<z<1.5$ & 10.64 & -3.14 & -1.37 \\
         $1.5<z<2.0$ & 11.01 & -4.05 & -1.74 \\ 
         $2.0<z<2.5$ & 10.93 & -3.93 & -1.77 \\
         $2.5<z<3.0$ & 11.08 & -4.41 & -1.92 \\ 
         \hline
         \hline
    \end{tabular}
    \label{tab:schetcher}
\end{table}

\begin{figure}
\begin{center}
\includegraphics[width=\linewidth]{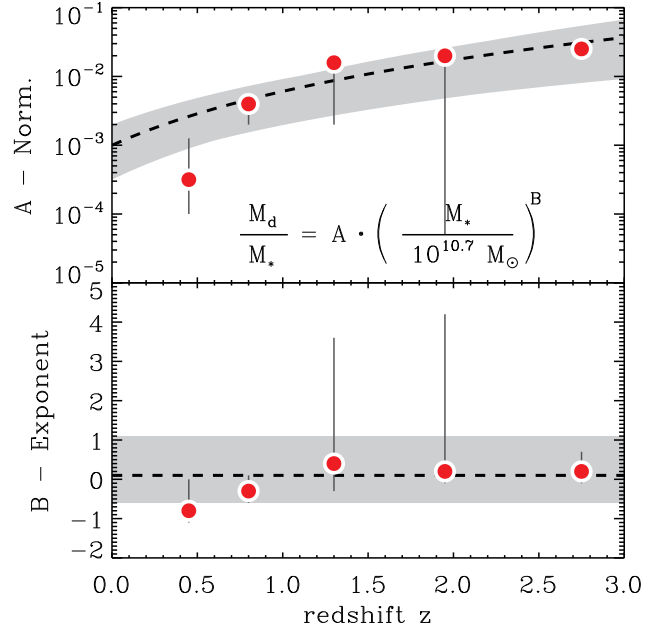}	
\caption{\label{fig:fdust} 
        Redshift evolution of the normalization (top panel) and exponent (bottom panel) of the dust-to-stellar mass ratio of star-forming galaxies modelled as a simple power-law of the stellar mass (see inset equation).
        $A$ is the typical dust-to-stellar mass ratio of star-forming galaxies with a stellar mass of $5\times10^{10}M_{\odot}$.
        Red circles show our constraints while fitting $\rho_{\rm dust}(M_\ast>M,z)$ in each redshift bin independently using Equation~\ref{eq:rho_dust_model}.
        Dashed lines show our constraints while fitting all redshift bins simultaneously and modelling the dust-to-stellar mass ratio as a simple power-law of the stellar mass, with a redshift-independent exponent and a redshift-dependent normalization (see Equation~\ref{eq:fdust_tacconi}). 
        Gray regions present the range of fits compatible within 1$\sigma$ with our observations.
        }
\end{center}
\end{figure}

First, we fitted each redshift bin independently, assuming that $f_{\rm dust}(M,z)$ follows a simple power-law,
\begin{equation}
    f_{\rm dust}(M,z)=A\times \bigg(\frac{M}{10^{10.7}\,M_{\odot}}\bigg)^{B},
\end{equation}
where the choice of a $10^{10.7}\,M_{\odot}$ normalization allows direct comparisons with \citet{tacconi_2018}.
With this parametrization, $A$ has an immediate physical interpretation, i.e., it is the typical dust-to-stellar mass ratios of star-forming galaxies with a stellar mass of $5\times10^{10}M_{\odot}$.
The results of these fits are shown by the black and light gray lines in Figure~\ref{fig:Mdust_vs_Mstar}, while the redshift evolution of $A$ and $B$ are shown by the red circles in Figure~\ref{fig:fdust}.
In each redshift bin, our fit provides an accurate description of our observations, with reduced $\chi^2$ in the range 0.5 to 1.0.
The exponent of the dust-to-stellar mass ratio is found to be negative ($B<0$) in the first two redshift bins but becomes positive at higher redshifts.
However, the uncertainties associated with these exponents render all of them consistent with zero at all redshifts.

\begin{figure*}
\begin{center}
\includegraphics[width=\linewidth]{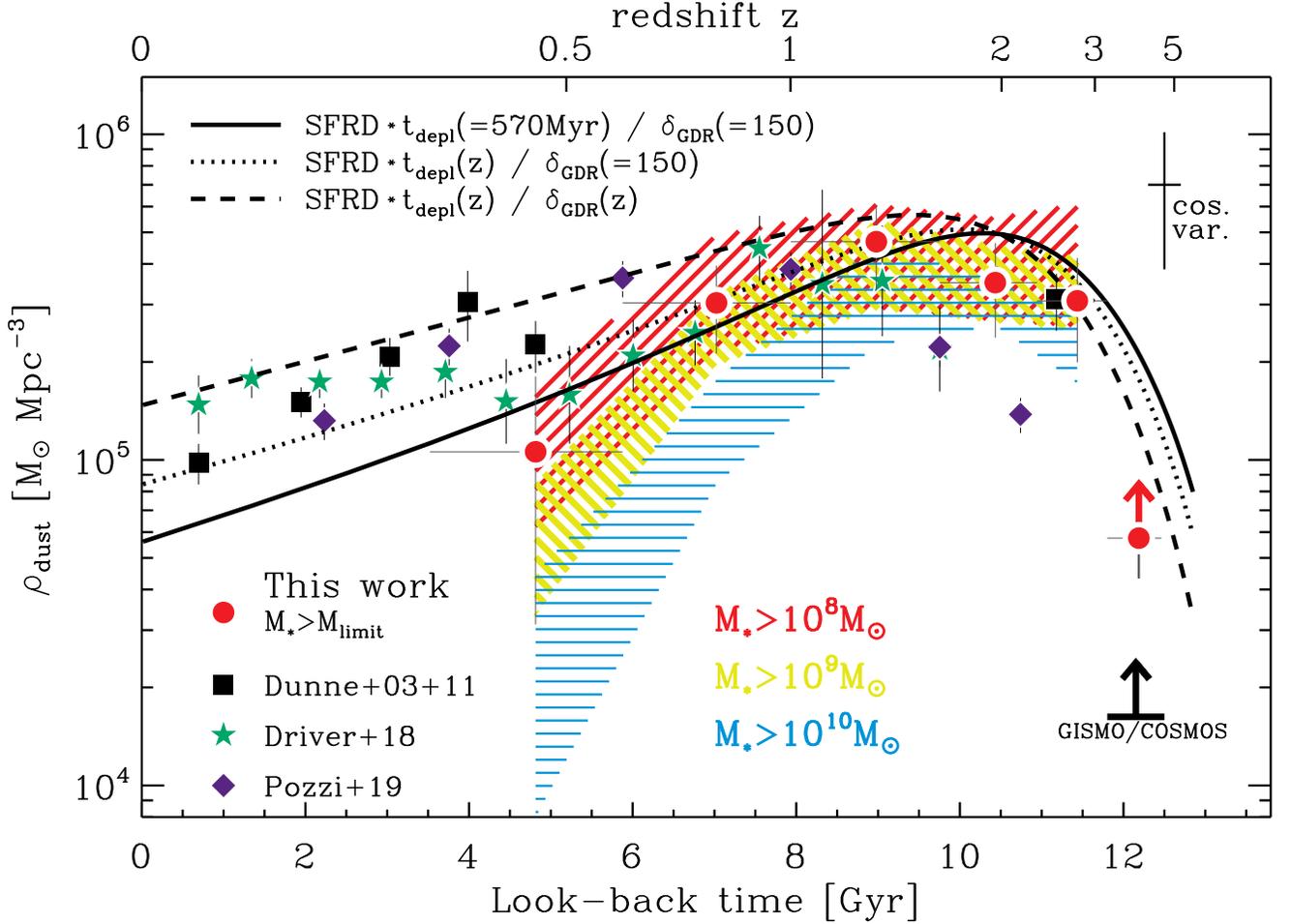}	
\caption{\label{fig:rho_dust_vs_redshift}
        Evolution with look--back time (lower $x$-axis; or redshift, upper $x$-axis) of the comoving dust mass density in galaxies derived here using the ASPECS LP 1.2\,mm continuum map.
        Red circles are for galaxies with $M_{\ast}>M_{\rm limit}$, as inferred from Figure~\ref{fig:Mdust_vs_Mstar}.
        Red, yellow and blue hashed regions are inferred from the best-fits of $\rho_{\rm dust}(M_\ast>M,z)$ (see Figure~\ref{fig:Mdust_vs_Mstar}) and correspond to the cosmic dust mass densities in galaxies with $M_{\ast}>10^{8}\,M_{\odot}$, $M_{\ast}>10^{9}\,M_{\odot}$, and $M_{\ast}>10^{10}\,M_{\odot}$, respectively.
        These best-fits were performed independently for each redshift bin.
        At $z>3$, we derive lower limits by stacking all galaxies in our $H$--band--selected sample, irrespective of their stellar masses. 
        At a similar redshift, we show the lower limit inferred by \citet{magnelli_2019} using a 2\,mm-selected galaxy sample.
        Black squares, green stars, and dark blue diamonds show the \textit{total} comoving dust mass densities in galaxies inferred by \citet{dunne_2003,dunne_2011}, \citet{driver_2018}, and \citet{pozzi_2019}, respectively.
        Solid line show predictions from scaling the cosmic SFRD \citep[]{madau_2014} assuming a redshift independent gas depletion time scale ($t_{\rm depl}=M_{\rm gas}/$SFR) of 570\,Myr and a gas-to-dust mass ratio of 150; both values being typical for $10^{10.3}\,M_\odot$ main-sequence star forming galaxies at $z\sim2$ \citep{tacconi_2018}.
        Dotted line show predictions assuming a gas-to-dust mass ratio of 150 but a redshift dependent gas depletion time as parametrized by \citet{tacconi_2018} for $10^{10.3}\,M_\odot$ main-sequence star forming galaxies (i.e., $t_{\rm depl}=570$\,Myr at $z=2$ and 860\,Myr at $z=0$).
        Finally, the dashed line show predictions assuming both a redshift dependent depletion time and gas-to-dust mass ratio. 
        This latter is derived from Equation~\ref{eq:gas_dust_ratio} using the metallicity of $10^{10.3}\,M_\odot$ galaxies at a given redshift (Equation~\ref{eq:metallicity}). 
        This yields $\delta_{\rm GDR}=150$ at $z=2$ and $\delta_{\rm GDR}=90$ at $z=0$.
        The typical fractional cosmic variance uncertainty of $\sim45\%$ affecting our measurements and inferred using the methodology advocated by \citet{driver_2010}, is illustrated in the upper right corner.}
\end{center}
\vspace{0.2cm}
\end{figure*}

We then solved for $f_{\rm dust}(M,z)$ by fitting $\rho_{\rm dust}(M_\ast>M,z)$ but fitting all redshift bins simultaneously and assuming that the exponent does not vary with redshift.
For the redshift-dependent normalization, we used a parametrization suggested by \citet{tacconi_2018}, i.e.,
\begin{equation}
\label{eq:fdust_tacconi}
\begin{array}{l}
   {\rm log}(f_{\rm dust}(M,z)) = C + D \times {\rm log}(1+z)\ + \\
   \hspace{2.9cm} B \times {\rm log}(M/(10^{10.7}\,M_{\odot})).
   \vspace{0.29cm}
\end{array}
\end{equation}
The best-fit is shown by the black dashed lines in Figure~\ref{fig:Mdust_vs_Mstar}, while the corresponding normalization (i.e., ${\rm log}(A) = C + D \times {\rm log}(1+z)$; with $C=-3.0^{+0.3}_{-0.5}$, and $D=2.6^{+1.1}_{-1.1}$) and exponent (i.e., $B=0.1^{+1.0}_{-0.7}$) of the dust-to-stellar mass ratio are shown in Figure~\ref{fig:fdust}.
This fit described accurately our observations, with a reduced $\chi^2=1.0$.
The mean dust-to-stellar mass ratio in star-forming galaxies is found to significantly vary with redshift.
Taking into account all models compatible, within 1$\sigma$ with our observations, we found that the mean dust-to-stellar mass ratio of $10^{10.7}\,M_{\odot}$ star-forming galaxy increases by at least a factor of 2 and at most a factor of 60 between $z=0.45$ and $2.75$.
Our best-fit implies an increase by a factor 10, which should be compared to the factor $12$ increase of the gas-to-stellar mass ratio found by \citet{tacconi_2018} within this redshift range.

Finally, from the gas-to-stellar mass ratio of star-forming galaxies derived in \citet{tacconi_2018}, we predicted $\rho_{\rm dust}(M_\ast>M,z)$ at $M=10^{10}\,M_\odot$, assuming a gas-to-dust mass ratio of 100, typical for massive galaxies at $z\sim0$ \citep{leroy_2011}.
These predictions are in good agreement, within the total uncertainties, with our observations, but in our lowest redshift bin (see blue diamonds in Figure~\ref{fig:Mdust_vs_Mstar}).
Thus, even though at high stellar masses our analysis is affected by the low number of galaxies available within our pencil-beam survey (i.e., cosmic variance), our measurements agree with those inferred using larger, representative samples of massive star-forming galaxies. 

\subsection{$\rho_{\rm dust}$ vs. redshift}
\label{subsec:rho_dust_redshift}

The redshift evolution of $\rho_{\rm dust}(M_\ast>M,z)$ at our stellar mass-completeness limits is shown in Figure~\ref{fig:rho_dust_vs_redshift}, i.e., $\rho_{\rm dust}(M_\ast>M_{\rm limit},z)$.
Along with these measurements, we also included estimates for $M_{\ast}>10^8\,M_{\odot}$, $M_{\ast}>10^9\,M_{\odot}$, and $M_{\ast}>10^{10}\,M_{\odot}$, inferred by fitting $\rho_{\rm dust}(M_\ast>M,z)$ in each redshift bin independently using the method described in Section~\ref{subsec:fdust}.
As advocated in Section~\ref{subsec:rho_dust_mass}, the \textit{total} cosmic dust mass density in galaxies (i.e., $\rho_{\rm dust}$) should be well approximated by these $M_{\ast}>10^8\,M_{\odot}$ measurements.

Our analysis suggests that $\rho_{\rm dust}$ did not evolve much from $z=2.75$ to $1.3$ but decreased by a factor $\sim3.6\,(\pm2.0)$ from $z=1.3$ to $0.45$.
As noted in Section~\ref{subsec:rho_dust_mass} and Figure~\ref{fig:Mdust_vs_Mstar}, this redshift evolution is consistent with that inferred by \citet{dunne_2003,dunne_2011} and \citet{driver_2018}.
It also broadly agrees with recent measurements by \citet{pozzi_2019} obtained using \textit{Herschel} observations in the COSMOS field. 
We only notice a significant disagreement with this later study at $z>2$, i.e., a redshift range where their observations mostly constrain the bright-end slope of the dust mass function. 

From $z=0.45$ to the present time, $\rho_{\rm dust}$ did not evolve significantly, as our $z=0.45$ measurement is already equal to the local ($z=0.05$) cosmic dust mass density in galaxies constrained by \citet{dunne_2011}.
We note, however, that these $z<0.45$ measurements rely mostly on \textit{Herschel}-250$\,\mu$m observations that probe the dust peak emission of galaxies, while our measurements rely on their dust Rayleigh-Jeans emission.
These two approaches might thus be affected by different systematics, which renders their combination difficult.

From early cosmic time to $z\sim3$, simulations predict a drastic increase of $\rho_{\rm dust}$ (see Section~\ref{subsec:simulations}) but further investigations are needed to confirm this trend observationally, e.g., using even longer wavelength selected samples (i.e., $\lambda_{\rm obs}>2\,\mu$m) which are still sensitive to stellar masses at these redshifts.

The redshift evolution of $\rho_{\rm dust}$ resembles that of the star-formation rate density ($\rho_{\rm SFR}$) of the Universe \citep[e.g.,][]{madau_2014}.
To investigate this further, we show in Figure~\ref{fig:rho_dust_vs_redshift} the redshift evolution of $\rho_{\rm SFR}$ \citep[][]{madau_2014} scaled assuming a (molecular) gas depletion time ($t_{\rm depl}=M_{\rm gas}/$SFR) of 570\,Myr and a gas-to-dust mass ratio ($\delta_{\rm GDR}$) of 150, i.e., 
\begin{equation}
    \rho_{\rm dust} = \rho_{\rm SFR} \times t_{\rm depl} \times \delta_{\rm GDR}^{-1}.
    \label{eq:toy model}
\end{equation}
These particular values of $t_{\rm depl}$ and $\delta_{\rm GDR}$ were chosen as they are typical for star-forming galaxies at $z\sim2$ with a stellar mass of $10^{10.3}\,M_\odot$ \citep[for $t_{\rm depl}$,][ for $\delta_{\rm GDR}$, Equations~\ref{eq:gas_dust_ratio} and \ref{eq:metallicity}]{tacconi_2018}, i.e., the characteristic stellar mass where most of the dust in galaxies is locked (Section~\ref{subsec:rho_dust_mass}).
While these predictions describe reasonably well our measurements at $z>0.45$, they underestimate the observations at low redshifts.
A flatter evolution of $\rho_{\rm dust}$ is predicted and thus a better match to the low redshift measurements is obtained, when assuming a more realistic redshift dependent gas depletion time (dotted line in Figure~\ref{fig:rho_dust_vs_redshift}), parametrized using the results from \citet[]{tacconi_2018} for $10^{10.3}\,M_\odot$ star-forming galaxies (i.e., $t_{\rm depl}=570$\,Myr at $z=2$ and 860\,Myr at $z=0$).
Finally, assuming both a redshift dependent depletion time and gas-to-dust mass ratio (dashed line in Figure~\ref{fig:rho_dust_vs_redshift}), we predict an even flatter evolution of $\rho_{\rm dust}$, which matches reasonably well our measurements. 
This flatter evolution illustrates the fact that at fix stellar mass (here $10^{10.3}\,M_\odot$), the mean metallicity of galaxies increases from $z\sim2$ to $z\sim0$ (see Equation~\ref{eq:metallicity}), which implies that their mean gas-to-dust mass ratio decrease over this redshift range ($\delta_{\rm GDR}=150$ at $z=2$ and $\delta_{\rm GDR}=90$ at $z=0$).

The flat evolution of $\rho_{\rm dust}$ at $z<0.45$ could also in part be due to the increasing contribution of the atomic phase of the ISM as we approach $z=0$.
This increasing contribution is not taken into account by our toy model, because the values for $t_{\rm depl}$ in Equation~\ref{eq:toy model} were taken from \citet{tacconi_2018} and only include the molecular gas phase.
At low redshifts, a significant contribution of the dust in the atomic phase to the total observed dust mass was actually reported by \citet{tacconi_2018}, when comparing CO-based and dust-based gas mass estimates.

As a final remark, we note that our estimates of $\rho_{\rm dust}$ are affected by cosmic variance, which is only partly included in the Poissonian uncertainties.
Using the method described by \citet{driver_2010}, we estimate for our 2.27\,arcmin$^2$ survey a fractional cosmic variance uncertainty of 58\%, 49\%, 42\%, 43\% and 42\% at $0.3\leq z<0.6$, $0.6\leq z<1.0$, $1.0\leq z<1.6$, $1.6\leq z<2.3$, and $2.3\leq z<3.2$, respectively.
These uncertainties are not significantly larger than the total uncertainties quoted in Table~\ref{tab:rho} and displayed in Figure~\ref{fig:rho_dust_vs_redshift}.

\begin{figure*}[!t]
\begin{center}
\includegraphics[width=\linewidth]{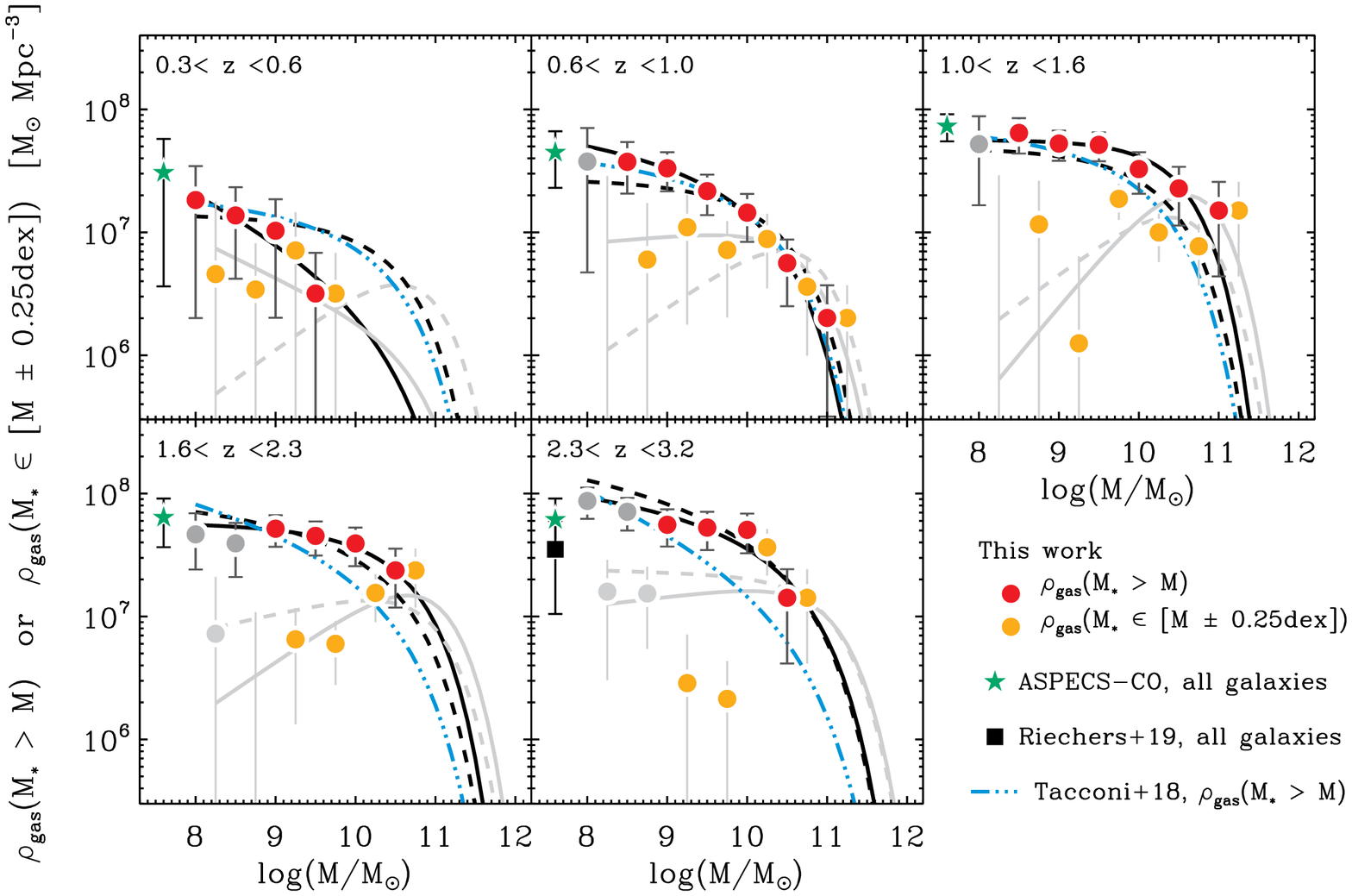}
\caption{\label{fig:Mgas_vs_mass}
        Same as Figure~\ref{fig:Mdust_vs_Mstar}, but for the comoving gas mass density in galaxies with stellar masses $>$$\,M$.
        Green stars present the \textit{total} comoving molecular gas mass density in galaxies measured by the ASPECS-CO pilot and LP surveys \citep{decarli_2016,decarli_2019}.
        The black square shows the \textit{total} comoving molecular gas mass density in galaxies measured by the COLDz survey \citep{riechers_2019}.
        Blue triple-dot-dashed lines show predictions using $f_{\rm gas}$($M$,$z$) for main-sequence galaxies from \citet{tacconi_2018} and the SMF of \citet{mortlock_2015}.
		}
\end{center}
\vspace{0.2cm}
\end{figure*}

\section{The cosmic gas mass density in galaxies}
\label{sec:rho_gas}

\subsection{$\rho_{\rm gas}(M_\ast>M,z)$ vs. $M_\ast$}
\label{subsec:rho_gas_mass}
The evolution of the comoving gas (H$_{2}+$\ion{H}{1}) mass density in galaxies with stellar masses $>M$,  i.e., $\rho_{\rm gas}(M_\ast>M,z)$ is shown in Figure~\ref{fig:Mgas_vs_mass} and tabulated in Table~\ref{tab:rho_gas}.
These measurements were inferred by stacking the 1.2\,mm emission of these galaxies (Section~\ref{subsec:stacking}) and assuming a metallicity-dependent gas-to-dust mass ratio relation (Section~\ref{subsec:measuring_mdust}).

As for the cosmic dust mass density, at $z>0.6$, $\rho_{\rm gas}(M_\ast>M,z)$ grows rapidly as $M$ decreases down to $\sim10^{10}\,M_{\odot}$, and this growth slows down as $M$ decreases to even lower stellar masses.
This flattening at low stellar masses suggests once again that at the stellar mass-completeness limits of our sample, the gas mass density measured here already accounts for most of the gas content locked in galaxies and converges toward the \textit{total} cosmic gas mass density in galaxies.
We note, however, that the slope of $\rho_{\rm gas}(M_\ast>M,z)$ at low stellar masses is slightly steeper than that inferred for $\rho_{\rm dust}(M_\ast>M,z)$.
This difference is explained by the decrease of the metallicity and therefore the increase of the gas-to-dust mass ratio at low stellar masses (Equation~\ref{eq:gas_dust_ratio} and \ref{eq:metallicity}).
Consequently, the peak of our differential measurements ($\rho_{\rm gas}(M_{\ast} \in [M\pm0.25 {\rm dex}],z)$) is much broader or somewhat washed out at $z<1.0$. 

Excluding quiescent galaxies from the stacking analysis barely affects our results, decreasing the value of $\rho_{\rm gas}(M_\ast>M,z)$ at our stellar mass-completeness limits by at most 10\%.
As in the case of dust, the bulk of the gas in galaxies appears to reside in star-forming galaxies \citep[see also][]{sargent_2015,gobat_2018}.

\citet{decarli_2016,decarli_2019} and \citet{riechers_2019} measured the \textit{total} molecular gas mass density in galaxies (i.e., $\rho_{\rm H_2}$) by constraining the CO luminosity function using the ASPECS-CO pilot/LP and COLDz surveys.
In all redshift bins, we find a good agreement between our measurements and their estimates.
This suggests that even though our gas mass measurements include in principle both the molecular and atomic gas phases, they are mostly dominated by the molecular phase.
We note, however, that our method implicitly assumes that the dust emissivity in the atomic and molecular gas phase is the same, while there exists observational evidence of an enhanced dust emissivity in the dense ISM \citep[see ][and reference therein]{leroy_2011}.
This implies that dust-based gas mass estimates might be biased against the dust in the atomic phase.

From the gas-to-stellar mass ratio of star-forming galaxies derived in \citet{tacconi_2018}, we predicted $\rho_{\rm gas}(M_\ast>M,z)$ using the SMF of \citet[]{mortlock_2015}.
Up to $z\sim2.0$, these predictions agree, within the uncertainties, with our observations, but in the highest redshift bin where they significantly underestimate our measurements (see triple-dot-dashed lines in Figure~\ref{fig:Mgas_vs_mass}).
At this redshift, the steep dependency with stellar mass of the gas-to-stellar mass ratio found in \citet[][$B=-0.36$]{tacconi_2018}, seem to over-estimate the contribution of low stellar mass galaxies to the cosmic gas mass density. 

Finally, we note that using a $\delta_{\rm GDR}$-metallicity relation with a much steeper power-law at metallicity $<7.9$ \citep[][]{remy_2014} only affects our estimates below our stellar mass-completeness limits (Figure~\ref{fig:Mgas_vs_mass_RR15} in Appendix~\ref{appendix:gas_to_dust}).
This steep power law, which implies much larger gas mass per unit dust mass at low metallicity, increases by a factor of 2 to 5 our measurements in our lowest stellar mass bins.
This leads to a very discontinuous evolution of $\rho_{\rm gas}(M_\ast>M,z)$ with $M$, large disagreements with the ASPECS-CO survey's results, and could suggest an increasing contribution of the atomic phase to $\rho_{\rm gas}(M_\ast>M,z)$ at low metallicities.
However, because the metallicity below which this steep power law starts is still very uncertain \citep{remy_2014}, we do not discuss this effect further.
\begin{figure}
\begin{center}
\includegraphics[width=\linewidth]{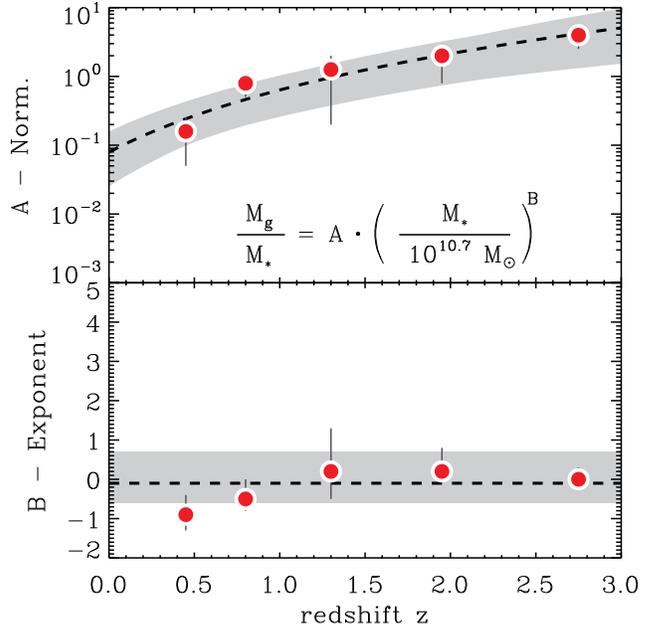}	
\caption{\label{fig:fgas} 
        Same as Figure~\ref{fig:fdust} but for the gas-to-stellar mass ratio of star-forming galaxies modelled as a simple power-law of the stellar mass (see inset equation).
        $A$ is the typical gas-to-stellar mass ratio of star-forming galaxies with a stellar mass of $5\times10^{10}M_{\odot}$.
        }
\end{center}
\end{figure}
\begin{figure*}
\begin{center}
\includegraphics[width=\linewidth]{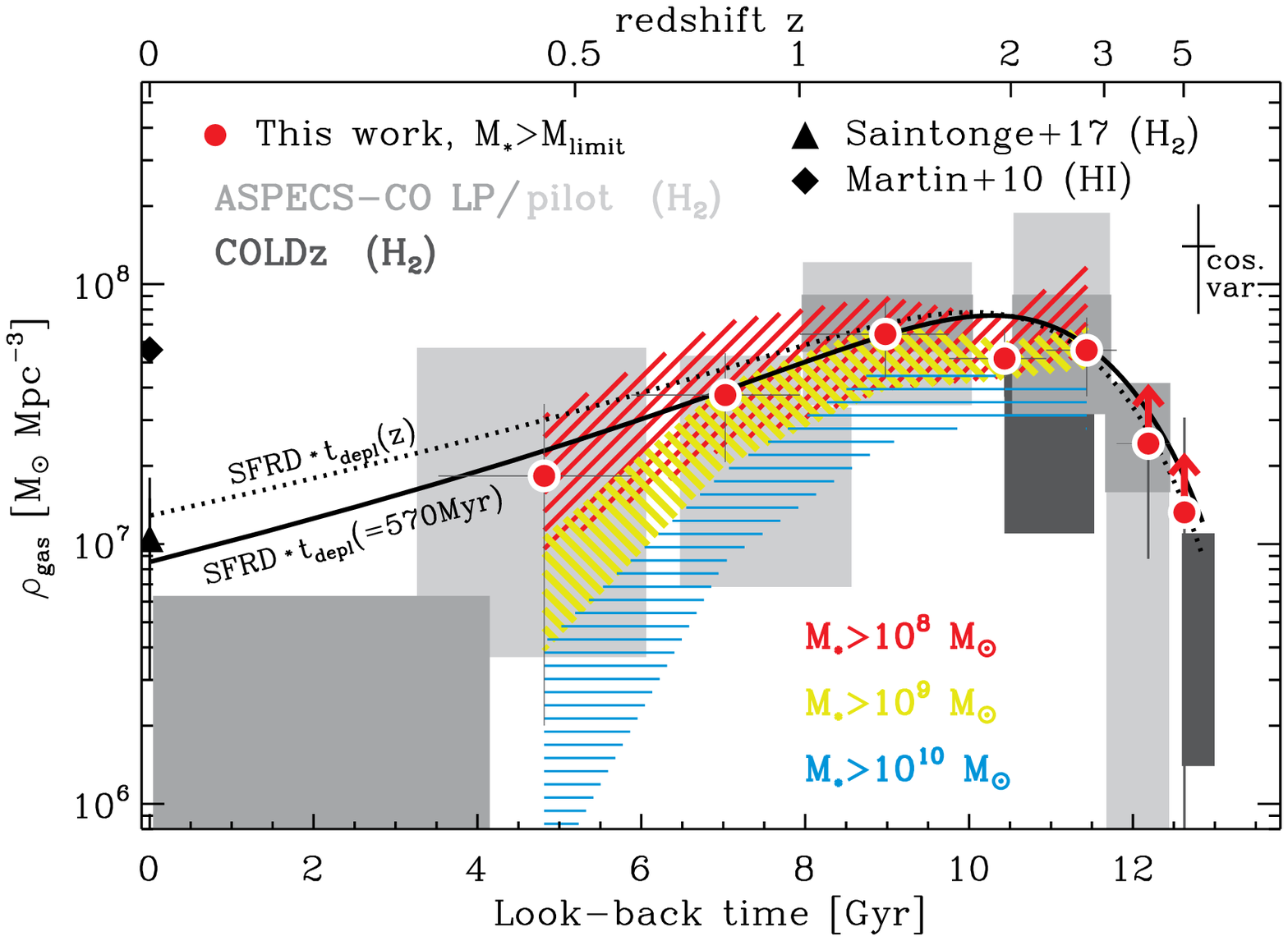}
\caption{\label{fig:rho_gas_vs_redshift}
        Same as Figure~\ref{fig:rho_dust_vs_redshift} but for the comoving gas mass density in galaxies.
        Shaded regions present the 1$\sigma$ confidence measurement for the molecular gas from the ASPECS-CO pilot and LP surveys \citep{decarli_2016,decarli_2019} as well as the 5th-to-95th percentile confidence interval from the COLDz survey \citep{riechers_2019}.
        At $z\sim0$, the black triangle and diamond show the cosmic molecular and atomic gas mass densities in galaxies inferred by \citet[]{saintonge_2017} and \citet[]{martin_2010}, respectively.
        Solid line show predictions from scaling the cosmic SFRD \citep[]{madau_2014} assuming a redshift independent gas depletion time ($t_{\rm depl}=M_{\rm gas}/$SFR) of 570\,Myr, typical for $10^{10.3}\,M_\odot$ main-sequence star forming galaxies at $z\sim2$ \citep{tacconi_2018}.
        Dotted line show predictions assuming a more realistic redshift dependent gas depletion time for $10^{10.3}\,M_\odot$ main-sequence star forming galaxies \citep[i.e., $t_{\rm depl}=570$\,Myr at $z=2$ and 860\,Myr at $z=0$;][]{tacconi_2018}.
		}
\end{center}
\vspace{0.5cm}
\end{figure*}

\subsection{The gas-to-stellar mass ratio of star-forming galaxies}
\label{subsec:fgas}
Using the method described in Section~\ref{subsec:fdust}, we model the gas-to-stellar mass ratio of star-forming galaxies as a simple power-law function of their stellar mass. 
First, we solve for the normalization and exponent of this power-law function in each redshift bin independently (thick black line in Figure~\ref{fig:Mgas_vs_mass} and red dots in Figure~\ref{fig:fgas}); and then, we fit all our redshift bins simultaneously assuming a redshift-independent exponent (thick dashed line in Figure~\ref{fig:Mgas_vs_mass} and gray regions in Figure~\ref{fig:fgas}; $B=-0.1^{+0.8}_{-0.5}$, $C=-1.1^{+0.3}_{-0.5}$, and $D=3.0^{+1.1}_{-1.0}$). 

The best-fit model obtained from fitting all redshift bins simultaneously, yields a exponent of $-0.1$.
This tentative trend, in which massive galaxies have lower gas mass content per unit stellar mass than lower mass galaxies, is, nevertheless, not as steep as that found in massive high-redshift galaxies \citep[$B\sim-0.36$; e.g.,][]{magdis_2012,genzel_2015,tacconi_2018}.
However, in the local Universe and considering only the molecular gas phase, \citet[]{saintonge_2017} found a flatter evolution with stellar mass of the gas-to-stellar mass ratio at $<10^{10.2}\,M_\odot$. 
Finally, even though we constrained independently the $f_{\rm dust}$-$M_\ast$ (Section~\ref{subsec:fdust}) and $f_{\rm gas}$-$M_\ast$ relations, these are linked via the gas-to-dust mass ratio (Equation~\ref{eq:gas_dust_ratio}) and stellar mass-metallicity (Equation~\ref{eq:metallicity}) relations. 
Combining Equations~\ref{eq:gas_dust_ratio} and \ref{eq:metallicity}, one can predict that the exponent of the $f_{\rm dust}$-$M_\ast$ relation should be higher by 0.15--0.3 to that of the $f_{\rm gas}$-$M_\ast$ relation. 

The mean gas-to-stellar mass ratio in $10^{10.7}\,M_{\odot}$ star-forming galaxies increases by at least a factor of 2, at most a factor of 70, and for our best-fit a factor of 17 between $z=2.75$ and $0.45$.
These values should be compared to the factor 12 increase found by \citet{tacconi_2018} within this redshift range.

\subsection{$\rho_{\rm gas}$ vs. redshift}
\label{subsec:rho_gas_redshift}
Figure~\ref{fig:rho_gas_vs_redshift} presents the redshift evolution of $\rho_{\rm gas}$ at our stellar mass-completeness limits, as well as our extrapolations for galaxies with $M_{\ast}>10^8\,M_{\odot}$, $M_{\ast}>10^9\,M_{\odot}$, and $M_{\ast}>10^{10}\,M_{\odot}$, obtained by fitting $\rho_{\rm gas}(M_\ast>M,z)$ in each redshift bin independently using the method described in Section~\ref{subsec:fgas}.
Because $\rho_{\rm gas}(M_\ast>M,z)$ clearly flattens toward low stellar masses, extrapolations for $M_{\ast}>10^8\,M_{\odot}$ galaxies, should provide a good measurement of the \textit{total} cosmic gas mass density in galaxies.

The cosmic gas mass density in galaxies decreases by a factor $\sim1.6\,(\pm0.7)$ from $z=2.75$ and $1.3$ and then decreases by a factor $\sim2.8\,(\pm1.7)$ between $z=1.3$ and $0.45$.
This redshift evolution is consistent with that inferred using the ASPECS-CO measurements \citep{decarli_2016,decarli_2019} and the COLDz survey \citep{riechers_2019}.
At $z<0.45$, the decrease of the cosmic gas mass density seems to continue, when considering the molecular gas mass density measured at $z\sim0$ by \citet{saintonge_2017}.
However, considering instead the atomic gas mass density measured at $z\sim0$ by \citet{martin_2010} yields an opposite trend, illustrating the increased contribution of the atomic phase in the ISM of galaxies.
Finally, combined with the ASPECS-CO constraints at $z>3$, our measurements suggest a rapid increase of $\rho_{\rm gas}$ from $z=4$ to $2.75$.

The evolution of the (molecular) gas mass densities of galaxies from $z\sim4$ to $z\sim0$ resembles that of $\rho_{\rm SFR}$.
As in Section~\ref{subsec:rho_dust_redshift}, to study this further we plot in Figure~\ref{fig:rho_gas_vs_redshift} the redshift evolution of $\rho_{\rm SFR}$ scaled assuming (i) a redshift-independent gas depletion time of 570\,Myr (solid line) and (ii) a more realistic redshift-dependent gas depletion time for $10^{10.3}\,M_\odot$ star-forming galaxies \citep[dotted line; i.e., $t_{\rm depl}=570$\,Myr at $z=2$ and 860\,Myr at $z=0$;][]{tacconi_2018}.
Both predictions match strikingly well the observations from $z\sim4$ to $z\sim0$.
This finding strongly suggests that (i) at $z>0.45$, dust-based gas mass measurements are mostly dominated by the dust in the molecular phase, and (ii) the redshift evolution of the cosmic SFRD is mostly explained by the evolution of the molecular gas reservoir of galaxies (solid line), with variations of their star formation efficiency playing a secondary role (dotted line).

The evolution of $\rho_{\rm dust}$ and $\rho_{\rm gas}$ measured in our study are linked via the gas-to-dust mass ratio and stellar mass-metallicity relations.
At a given redshift, the $\rho_{\rm gas}$-to-$\rho_{\rm dust}$ ratio correspond thus to the mass-weighted average gas-to-dust mass ratio of star-forming galaxies. 
At $z\sim0.45$, $z\sim0.80$, $z\sim1.30$, $z\sim1.95$, and $z\sim2.75$, we found $166\pm115$, $155\pm104$, $130\pm73$, $140\pm63$, and $240\pm140$, respectively.
We note that in Figure~\ref{fig:rho_gas_vs_redshift_fix_gdr} of Appendix~\ref{appendix:gas_to_dust}, we infer instead the redshift evolution of $\rho_{\rm gas}$ scaling $\rho_{\rm dust}$ by a constant gas-to-dust mass ratio of 100, typical for massive galaxies at $z\sim0$.
These measurements underestimate at the $1-2\sigma$ level that from the ASPECS-CO LP at $z=1-2$.

%The gas-to-dust mass ratio (Equation~\ref{eq:gas_dust_ratio}) and stellar mass-metallicity (Equation~\ref{eq:metallicity}) relations used here are not yet well constrained up to the redshift and down to the stellar mass explored in this study. 
%While the agreement between our $\rho_{\rm gas}$ measurements and that from CO seems to validate our approach, these limitations should be kept in mind when using our $\rho_{\rm gas}$ measurements.
%In Figure~\ref{fig:rho_gas_vs_redshift_fix_gdr} of Appendix~\ref{appendix:gas_to_dust}, we infer the redshift evolution of $\rho_{\rm gas}$, simply scaling $\rho_{\rm dust}$ using a constant gas-to-dust mass ratio of 100, typical for massive galaxies at $z\sim0$.
%These measurements fall below that from the ASPECS-CO LP at $z=1-2$, yet consistent within 1--2$\sigma$.

\section{Discussion}
\label{sec:discussion}
Our results reveal that up to $z\sim3$ most of the dust and gas content in galaxies resides in star-forming galaxies with stellar masses $\gtrsim10^{9.5}\,M_\odot$.
Our cosmic gas mass density estimates agree with those inferred from CO observations, which suggests that dust-based measurements are dominated by the dust in the molecular phase.
The \textit{total} dust and gas mass densities in galaxies increase at early cosmic time, peak around $z\sim1-3$, and then decrease until the present time.
The dust and gas mass densities decrease, however, at a different rate; the former declines by a factor $\sim4$ between $z\sim2.5$ and 0 \citep[combining our results with the \textit{Herschel}-based local measurement of][]{dunne_2011}; while the later declines by a factor $\sim7$ when only considering the molecular gas phase and using the local CO measurement of \citet{saintonge_2017}.
The redshift evolution of the cosmic dust and gas mass densities can be modelled by the redshift evolution of the stellar mass function of star-forming galaxies and that of their dust-to-stellar mass ratio and gas-to-stellar mass ratio, respectively.
These models show that the dust and gas content of star-forming galaxies per unit stellar mass continuously decrease from $z=3$ to $z=0$, while having a mild dependency on stellar masses --with a best-fit power law stellar mass-dependent exponent of $0.1$ and $-0.1$ for the dust-to-stellar mass ratio and gas-to-stellar mass ratio relations, respectively.

In the following, we first compare these new results to the outputs of simulations and then put them into the context of galaxy evolution scenarios.

\subsection{Comparison to simulations}
\label{subsec:simulations}
\begin{figure*}
\begin{center}
\includegraphics[width=0.85\linewidth]{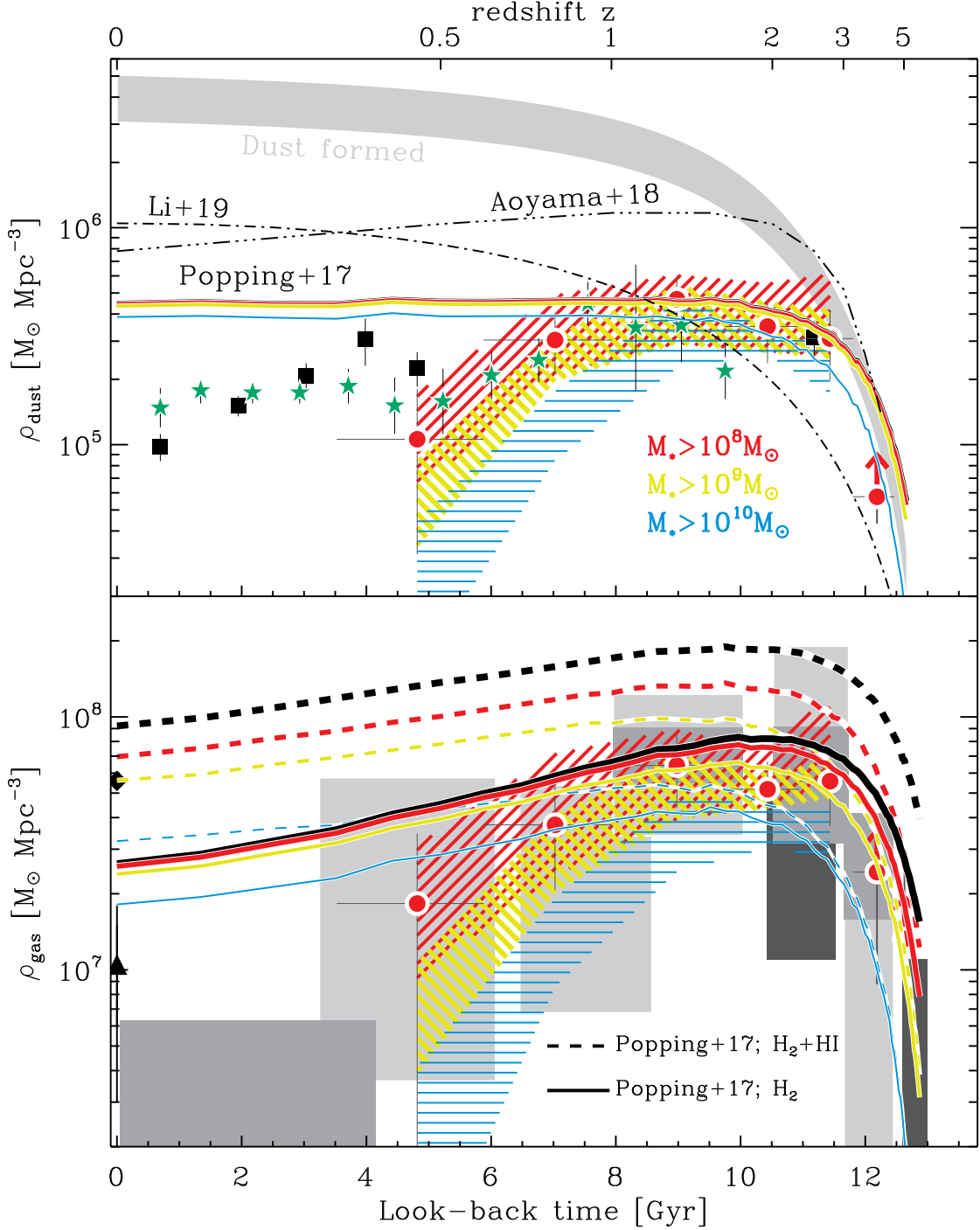}	
\caption{\label{fig:rho_vs_model}
        Evolution with look--back time of the observed comoving dust (top panel) and gas (bottom panel) mass densities in galaxies, compared to predictions from the semi-analytical model of \citet{popping_2017}.
        For the observed comoving dust and gas mass densities, symbols are the same as in Figures~\ref{fig:rho_dust_vs_redshift} and \ref{fig:rho_gas_vs_redshift}, respectively.
        (top panel) Blue, yellow, red and black solid lines show predictions for galaxies with $M_{\ast}>10^{10}\,M_{\odot}$, $>10^{9}\,M_{\odot}$, $>10^{8}\,M_{\odot}$, and $>10^{7}\,M_{\odot}$, respectively.
        The dot-dashed and triple-dot-dashed lines show predictions from the cosmic dust mass densities in galaxies from the cosmological hydrodynamic simulations of \citet[][see also Dav\'e et al. \citeyear{dave_2019}]{li_2019} and \citet{aoyama_2018}.
        The gray shaded area shows the total amount of dust formed in galaxies, assuming a dust yield of $0.004-0.0065$ dust masses for every unit of stellar mass formed \citep[][]{driver_2018} and using the cosmic SFRD history of \citet[][]{madau_2014}.
        (bottom panel) The dashed and solid lines show predictions from \citet[][]{popping_2017} for the gas (H$_{2}+$\ion{H}{1}) and molecular gas mass densities in galaxies, respectively.
        Lines are color-coded as in the top panel.
		}
\end{center}
\vspace{0.2cm}
\end{figure*}
In the past years, there has been a growing interest in including self-consistent tracking of the production and destruction of dust in cosmological models of galaxy formation \citep[e.g.,][]{popping_2017,mckinnon_2017,aoyama_2018,dave_2019,vijayan_2019}.
This includes the condensation of dust in the ejecta of asymptotic giant branch (AGB) stars and supernovae (SNe), the growth of dust in the ISM, as well as the destruction of dust by supernova-induced shocks, star formation (the so-called astraction), reheating, and outflows.
Our analysis provides to these models some of the first robust and homogeneously constrained evolution of the comoving dust and gas mass densities in galaxies from $z\sim3$ to $z\sim0.4$. 

In Figure~\ref{fig:rho_vs_model}, we compare the cosmic dust and gas mass densities predicted by the semi-analytical model of \citet[][]{popping_2017} for different stellar mass thresholds. 
These predictions correspond to the dust and gas in the ISM of these simulated galaxies, as opposed to the dust and gas in their circumgalactic medium (CGM) and the intergalactic medium \citep[IGM; see][for details]{popping_2017}.
In agreement with our observations, this model predicts that at any redshifts the bulk of the dust in galaxies is locked in those with a stellar mass $\gtrsim10^{10}\,M_{\odot}$.
However, while these predictions match the observations from $z\sim3$ to $z\sim1$, they over-estimate the cosmic dust mass density in galaxies at $z<1$.
In this model, from $z\sim1$ to $z\sim0$, the formation and destruction (or ejection) rates of the dust are in quasi-equilibrium, yielding a nearly constant cosmic dust mass density across this redshift range.
Instead, observations suggest that such quasi-equilibrium is reached at later time, i.e., $z<0.5$.
We note, however, that to confirm this trend, one would need to perform across this redshift range a more homogeneous observational analysis.
Indeed, low redshift measurements rely mostly on 250$\,\mu$m \textit{Herschel} observations that probe the dust peak emission of galaxies, while our measurements rely on their dust Rayleigh-Jeans emission.
These two approaches might thus be affected by different systematics, which renders their combination difficult.

In the top panel of Figure~\ref{fig:rho_vs_model}, we also overlaid the total cosmic dust mass densities in galaxies predicted by the cosmological hydrodynamic simulations of \citet{aoyama_2018} and \citet[][see also Dav\'e et al. \citeyear{dave_2019}]{li_2019}.
Those are relatively inconsistent with our measurements, over-estimating at most redshifts the cosmic dust mass densities in galaxies.

In the bottom panel of Figure~\ref{fig:rho_vs_model}, we finally compare the cosmic gas (H$_{2}+$\ion{H}{1}) and molecular gas mass densities predicted by \citet[][]{popping_2017} for different stellar mass thresholds.
Predictions for the molecular gas phase match reasonably well the observations from $z\sim3$ up to $z\sim0$.
In addition, they also correctly predict that most of the molecular gas mass is locked in galaxies with $\gtrsim10^{9}\,M_{\odot}$.
However, our dust-based measurements should in principle be compared to predictions including both the molecular and atomic gas phases.
In this case, predictions from \citet[][]{popping_2017} over-estimate our measurements by a factor 2 for galaxies with $M_\ast>10^{10}\,M_{\odot}$ and a factor 4 for galaxies with $M_\ast>10^{8}\,M_{\odot}$, illustrating the rising contribution in this model of the atomic gas phase in the ISM of low stellar mass galaxies.
These over-estimations suggest that either the model over-estimates the atomic gas content locked in low stellar mass galaxies or that our measurements are biased against this atomic gas phase because of a significant enhancement of the dust emissivity in the dense/molecular ISM.\\

In Figure~\ref{fig:vijayan} of Appendix~\ref{appendix:simulations}, we compare our measurements to predictions from the semi-analytical model of \citet[][]{vijayan_2019} for $M_\ast>10^8\,M_\odot$ and $M_\ast>10^9\,M_\odot$.
For the cosmic dust mass density in galaxies, these predictions are very similar to those of \citet{popping_2017}, i.e., this model successfully predicts little contribution to $\rho_{\rm dust}$ from $M_\ast<10^{9}\,M_{\odot}$ galaxies, but overestimates $\rho_{\rm dust}$ at $z<1$.
For the cosmic gas mass density in galaxies, predictions from \citet{vijayan_2019} differ from those of \citet{popping_2017}. 
They overestimate $\rho_{\rm gas}$ at $z\lesssim1.5$ but successfully suggest that at high redshift most of the gas in galaxies is in the molecular phase.

\subsection{Implications for galaxy evolution scenarios}
The redshift evolution of $\rho_{\rm dust}$ and $\rho_{\rm gas}$ resembles that of the cosmic SFRD, which also peaks at $z\sim2$ \citep[e.g.,][]{madau_2014}. 
This implies a direct link between star formation, and the dust and gas content of galaxies.
For the gas, such a link is expected --gas fueling star-formation-- via the so-called Kennicutt-Schmidt (KS) relation, which connects the gas and star-formation rate surface densities of galaxies.
The relatively redshift-independent SFRD-to-$\rho_{\rm gas}$ ratio observed here from $z\sim3$ to $z\sim0$ (see solid black line in Figure~\ref{fig:rho_gas_vs_redshift}) suggests that the global star-formation efficiency of galaxies (i.e., SFR/$M_{\rm gas}$) does not evolve significantly ($\lesssim\times2$) across this redshift range and that the main driver of star-formation is gas content.
At all redshifts, the time needed to deplete the global gas reservoir of the star-forming galaxy population ($\langle t_{\rm depl}\rangle_{\rm V}=\rho_{\rm gas}$/SFRD) is found to be of the order of 600--900\,Myr, consistent with results inferred from individual galaxies \citep{genzel_2015,tacconi_2018}.
Without a constant replenishment of these gas reservoirs, the star-forming galaxy population observed at, e.g., $z\sim2$, would thus have fully disappeared by $z=1.5$.

These findings strongly support `gas regulator models' \citep[e.g.,][]{bouche_2010,dave_2012,lilly_2013,peng_2014,rathaus_2016}, in which galaxy growth is mostly driven by a continuous supply of fresh gas from the cosmic web \citep[][]{dekel_2009}. 
In these models, gas accretion on halos and subsequently galaxies are controlled by the expansion of the Universe.
It decreases as $(1+z)^{2.3}$ and scales nearly linearly with halo masses \citep[e.g.][]{neistein_2008}.
This redshift evolution agrees with the continuous decline of the gas-to-stellar mass ratio of galaxies observed here from $z=3$ to 0.45.
In addition, `gas regulator models' generally invoke feedback processes --such as stellar winds-- to suppress or slow down gas accretion on low mass halos ($<10^{11}\,M_\odot$) \citep[e.g.][]{bouche_2010,dave_2012}. 
This introduces a mass-dependency of the gas-to-stellar mass ratio consistent with our observations (best-fit exponent of $-0.1$, though with large uncertainties) and delays the cosmic SFRD peak to $z\sim2$, because of the relatively low number of massive enough halos at early cosmic time \citep[][]{bouche_2010}. 
To first order, the rise of the SFRD and $\rho_{\rm gas}$ from cosmic dawn to $z\sim2$ would thus be due to the increased number of halos experiencing efficient gas accretion, which is observationally consistent with the increase of the stellar mass function of star-forming galaxies at these epochs.
The decrease of the SFRD and $\rho_{\rm gas}$ at $z<2$ would then be mostly controlled by the decrease of the gas accretion due to the Universe's expansion or due to shock heating preventing accretion on the most massive halo, which is observationally consistent with the decrease of the gas-to-stellar mass ratio in star-forming galaxies.\\

A link between the star-formation and dust content of galaxies is also expected. 
Stars produce the metals needed for the formation of dust and at the end of their life-cycle are the locus of significant dust formation, either via an AGB phase ($M\lesssim8M_{\odot}$) or SNe \citep[$M\gtrsim8M_{\odot}$; e.g.,][]{gall_2011,gall_2014}.
The coincident peak of the cosmic dust mass density and SFRD suggests a very important contribution of SNe and AGB stars to dust formation, as these formation pathways are linked to star-formation on a timescale of 1--2 Gyr \citep[e.g.,][]{dwek_2007,valiante_2009}.
However, ISM dust grain growth is supposedly also enhanced in high-redshift star-forming galaxies with high gas densities, and can thus contribute significantly, as well as on a short time scale, to the dust production in these galaxies \citep{popping_2017}. 
A significant contribution of this latter mechanism to the global dust content of galaxies cannot be ruled out from our observations.

As already noted by \citet{driver_2018}, the decrease of $\rho_{\rm dust}$ at $z<2$ is at odds with a close-box scenario in which the dust continuously accumulates in galaxies in the absence of destruction or expulsion mechanisms.
Our observations suggest instead that at $z<2$, the dust is destroyed (or expelled) more rapidly than it is formed.
Predicting the total amount of dust formed assuming a dust yield of 0.004--0.0065 dust masses for every unit of stellar mass formed as in \citet[][gray line in Figure~\ref{fig:rho_vs_model}; i.e., implicitly assuming an redshift-independent initial mass function]{driver_2018}, we infer that at $z\sim0$ about 90\% of the dust that has been formed in the Universe must be destroyed (e.g., astraction, supernovae shocks) or ejected in the IGM (e.g., stellar winds, radiation pressure).

Finally, the decrease of $\rho_{\rm dust}$ at $z<2$ is found to be not as pronounced as that of $\rho_{\rm gas}$, when considering only the molecular phase \citep[i.e., $\rho_{\rm H_2}(z\sim0)$ from ][]{saintonge_2017}.
This flatter evolution of $\rho_{\rm dust}$ can be explained by the increased contribution of the atomic phase of the ISM and the increased mean metallicity of star-forming galaxies as we approach $z=0$.
Indeed, even though the gas reservoirs of galaxies are replenished by pristine gas from the IGM, their overall metallicity increases from $z\sim2$ to $z\sim0$ --as confirmed by the increased zero point of the stellar mass-metallicity relation--, implying a lower global gas-to-dust mass ratio at $z=0$ than at $z=2$.

\section{Conclusions}
\label{sec:conclusion}
We used the deepest ALMA 1.2\,mm continuum map to date (rms: 9.5 $\mu$Jy\,beam$^{-1}$) in the HUDF obtained as part of the ASPECS large program to measure the cosmic dust and implied gas (H$_{2}+$\ion{H}{1}) mass densities as a function of look--back time.
We do this by summing (i.e., stacking) the contribution of all the known galaxies in the HUDF above a given stellar mass in distinct redshift bins, i.e., $\rho_{\rm dust}(M_\ast>M,z)$ and $\rho_{\rm gas}(M_\ast>M,z)$.
Our galaxy sample is $H$-band--selected from the available HUDF multi-wavelength catalogue, and, up to $z\sim3$, can be considered as stellar mass-complete down to $\sim10^{8.9}\,M_{\odot}$.
Dust masses are measured from the 1.2\,mm emission of these galaxies assuming a mass-weighted mean dust temperature of $\langle T\rangle_{\rm M}=25\,$K and a dust emissivity of $\beta=1.8$ \citep{scoville_2016}.
Gas masses are inferred from these dust masses using the local metallicity-dependent gas-to-dust mass ratio and the redshift-dependent stellar mass-metallicity relations \citep[as in][]{tacconi_2018}. 
With this unique dataset and approach, we find the following:
\begin{enumerate}
    \item $\rho_{\rm dust}(M_\ast>M,z)$ and $\rho_{\rm gas}(M_\ast>M,z)$ grow rapidly as $M$ decreases down to $10^{10}\,M_\odot$, but this growth slows down as $M$ decreases to even lower stellar masses. This flattening implies that at our stellar mass-completeness limits, $\rho_{\rm dust}(M_\ast>M,z)$ and $\rho_{\rm gas}(M_\ast>M,z)$ converge already towards the \textit{total} cosmic dust ($\rho_{\rm dust}$) and gas ($\rho_{\rm gas}$) mass densities in galaxies, i.e., with only a minor contribution by galaxies below our stellar mass-completeness limits. 
    \item The contribution of quiescent galaxies -- i.e., galaxies with little on-going star-formation and selected using the standard $UVJ$ criterion-- to $\rho_{\rm dust}$  and $\rho_{\rm gas}$ is negligible ($\lesssim10\%$). The bulk of the dust and gas in galaxies appears to be locked in star-forming galaxies with $M_{\ast}\gtrsim10^{9.5}\,M_\odot$.
    \item The gas (H$_{2}+$\ion{H}{1}) mass densities measured here agree with the molecular gas mass densities inferred from the CO observations of the ASPECS \citep{decarli_2016,decarli_2019} and COLDz \citep{riechers_2019} surveys. In the redshift range probed here ($z=0.45$ to $3.0$), dust-based measurements are thus dominated by the dust in the molecular phase. This suggests that either the bulk of the gas in galaxies is in a molecular phase or that there is a significant enhancement of the dust emissivity in the dense/molecular ISM with respect to that in the more diffuse/atomic ISM.
    \item The cosmic dust (gas) mass density increases at early cosmic time, peaks around $z=1-3$, and decreases by a factor $\sim4$ (7) until the present time \citep[combining our results with low redshift measurements;][]{dunne_2011,decarli_2016,saintonge_2017,decarli_2019}. The redshift evolution of $\rho_{\rm gas}$ matches that of the cosmic SFRD, while the decline of $\rho_{\rm dust}$ at $z<2$ is less pronounced than that observed for the SFRD.
    \item The evolution of $\rho_{\rm dust}(M_\ast>M,z)$ [$\rho_{\rm gas}(M_\ast>M,z)$] with stellar masses and redshifts is accurately modelled using the stellar mass function of star-forming galaxies and their average dust[gas]-to-stellar mass ratio. The dust [gas] content of galaxies per unit stellar mass continuously decreases from $z=3$ to $z=0$, while having a mild dependency on stellar masses --with a best-fit power-law stellar mass-dependent exponent of $0.0$ [$-0.1$].
\end{enumerate}

Our results suggest that galaxies have a relatively constant star-formation efficiency (SFR/$M_{\rm gas}$) across cosmic time (within a factor $\sim2$; solid line in Figure~\ref{fig:rho_gas_vs_redshift}).
Their star-formation seems mainly controlled by the supply of fresh gas from the cosmic web \citep{dekel_2009}, as advocated by the `gas regulator models' \citep[e.g.,][]{bouche_2010,dave_2012,lilly_2013,peng_2014,rathaus_2016}.
This supply, which varies with cosmic time as $(1+z)^{2.3}$ following the Universe's expansion, is in turn the main driver of the continuous decrease of the SFRD at $z<2$.
The decrease of $\rho_{\rm dust}$ at $z<2$ implies that a large fraction \citep[$\sim90\%$; see][]{driver_2018} of the dust formed in galaxies is, within few Gyr, destroyed (shock, astraction) or ejected to the IGM (wind, radiation pressure).

\acknowledgements
We would like to thank the referee for comments that have helped improve the paper.
This research was carried out within the Collaborative Research Centre 956, sub-project A1, funded by the Deutsche Forschungsgemeinschaft (DFG) -- project ID 184018867.
B.M. thanks Aswin P. Vijayan for providing predictions from the L-GALAXIES semi-analytic model.
M.N. and F.W. acknowledge support from ERC Advanced Grant 740246 (Cosmic\_Gas).
D.R. acknowledges support from the National Science Foundation under grant number AST-1614213 and from the Alexander von Humboldt Foundation through a Humboldt Research Fellowship for Experienced Researchers.
IRS acknowledges support from STFC (ST/P000541/1).
RJA was supported by FONDECYT grant number 1191124.
Este trabajo cont\'o con el apoyo de CONICYT + PCI + INSTITUTO MAX PLANCK DE ASTRONOMIA MPG190030.
Este trabajo cont\'o con el apoyo de CONICYT + Programa de Astronom\'ia+ Fondo CHINA-CONICYT CAS16026.
This Paper makes use of the ALMA data \newline ADS/JAO.ALMA\#2016.1.00324.L. ALMA is a partnership of ESO (representing its member states), NSF (USA) and NINS (Japan), together with NRC (Canada), NSC and ASIAA (Taiwan), and KASI (Republic of Korea), in cooperation with the Republic of Chile.
The Joint ALMA Observatory is operated by ESO, AUI/NRAO and NAOJ.
The National Radio Astronomy Observatory is a facility of the National Science Foundation operated under cooperative agreement by Associated Universities, Inc.
\vspace{5mm}

\facilities{ALMA}

\appendix

\section{Different Gas-to-Dust Mass Conversion}
\label{appendix:gas_to_dust}

\begin{figure*}
\begin{center}
\includegraphics[width=0.9\linewidth]{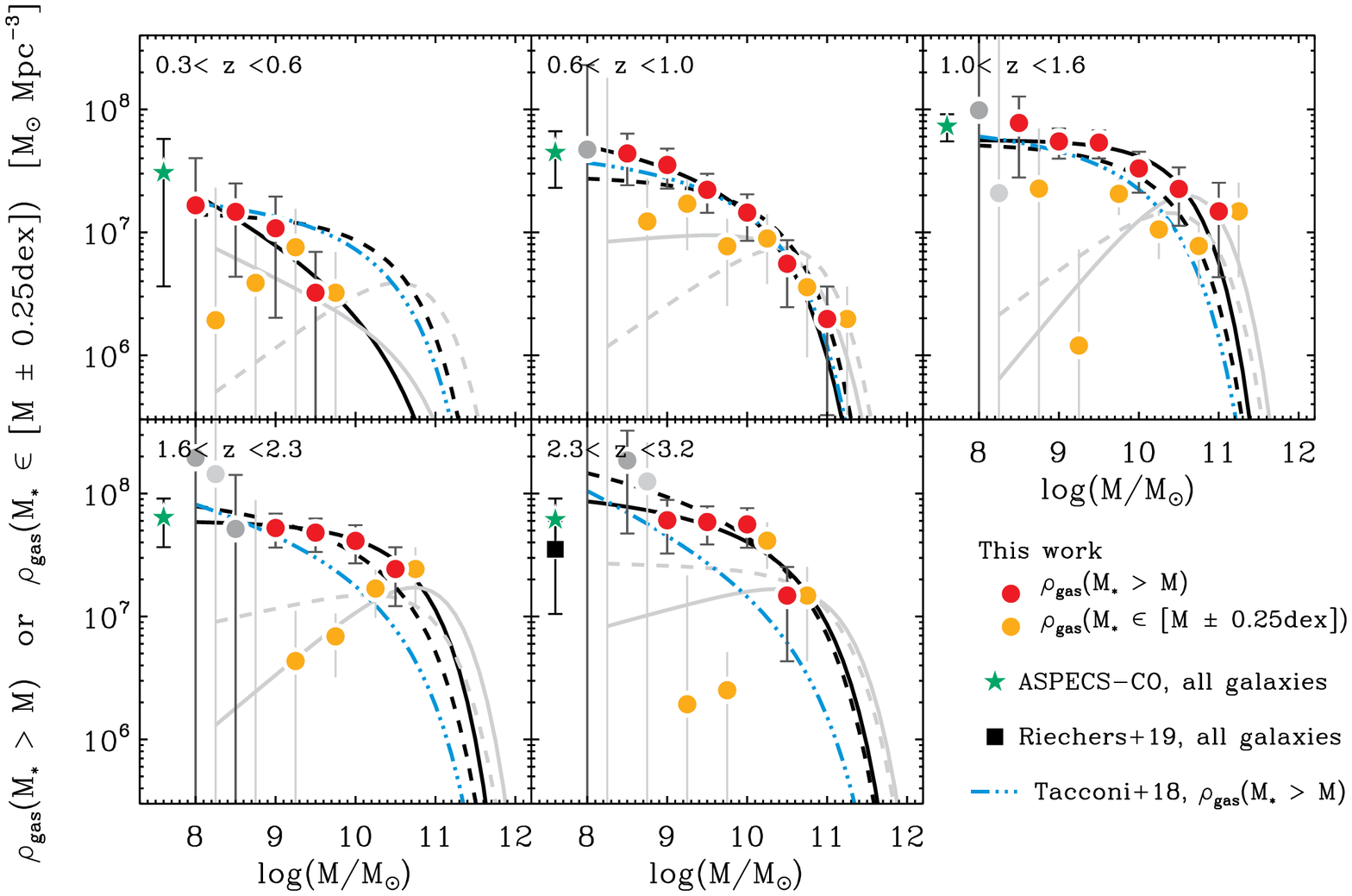}
\caption{\label{fig:Mgas_vs_mass_RR15}
        Same as Figure~\ref{fig:Mgas_vs_mass} but here the gas mass densities are inferred using a gas-to-dust mass ratio-metallicity relation with a much steeper power-law at metallicity $<7.9$ \citep[$\sim\,3$;][]{remy_2014}.
        }
\end{center}
\vspace{0.2cm}
\end{figure*}

\begin{figure*}
\begin{center}
\includegraphics[width=0.9\linewidth]{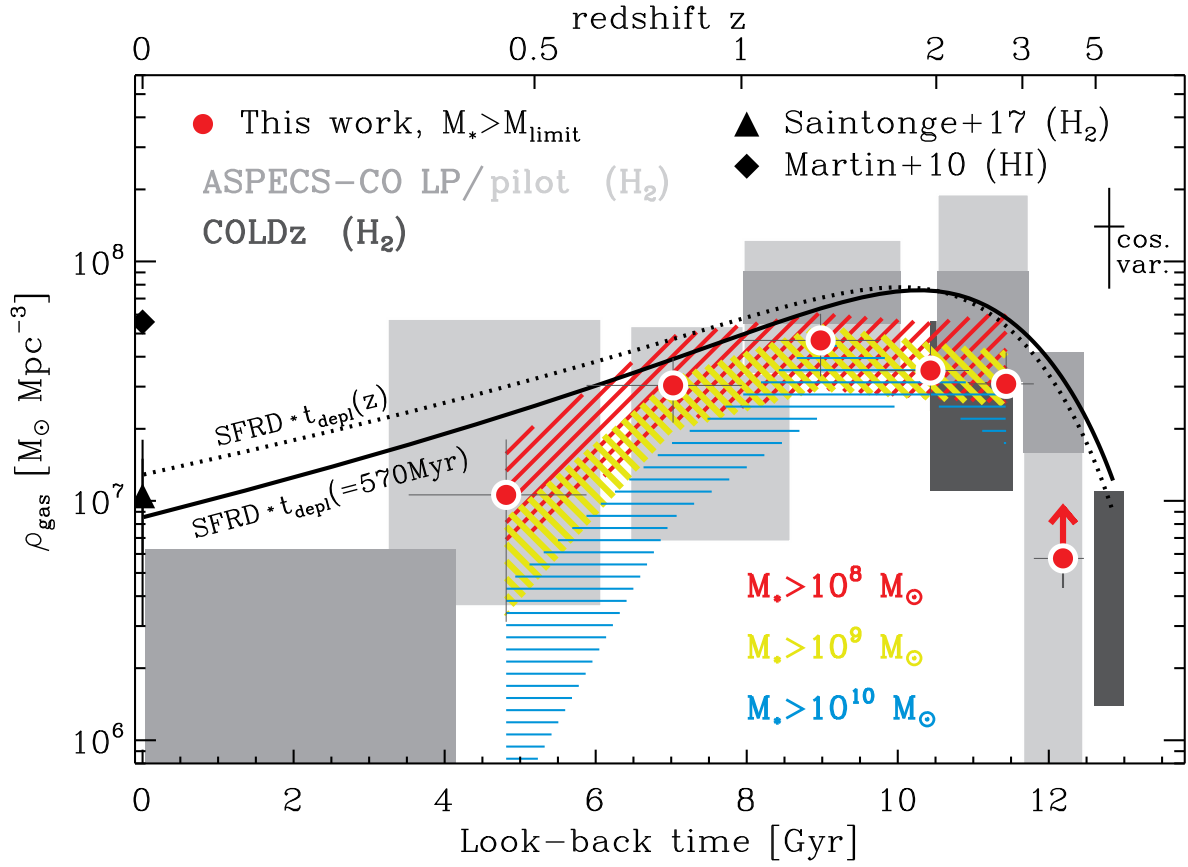}
\caption{\label{fig:rho_gas_vs_redshift_fix_gdr}
        Same as Figure~\ref{fig:rho_gas_vs_redshift} but here the cosmic gas mass densities are inferred applying a constant gas-to-dust mass ratio of 100 (typical for massive galaxies at $z\sim0$) to the cosmic dust mass densities presented in Figure~\ref{fig:rho_dust_vs_redshift}.
        }
\end{center}
\vspace{0.2cm}
\end{figure*}

\section{Comparison to other simulations}
\label{appendix:simulations}

\begin{figure*}
	\begin{center}
		\includegraphics[width=0.85\linewidth]{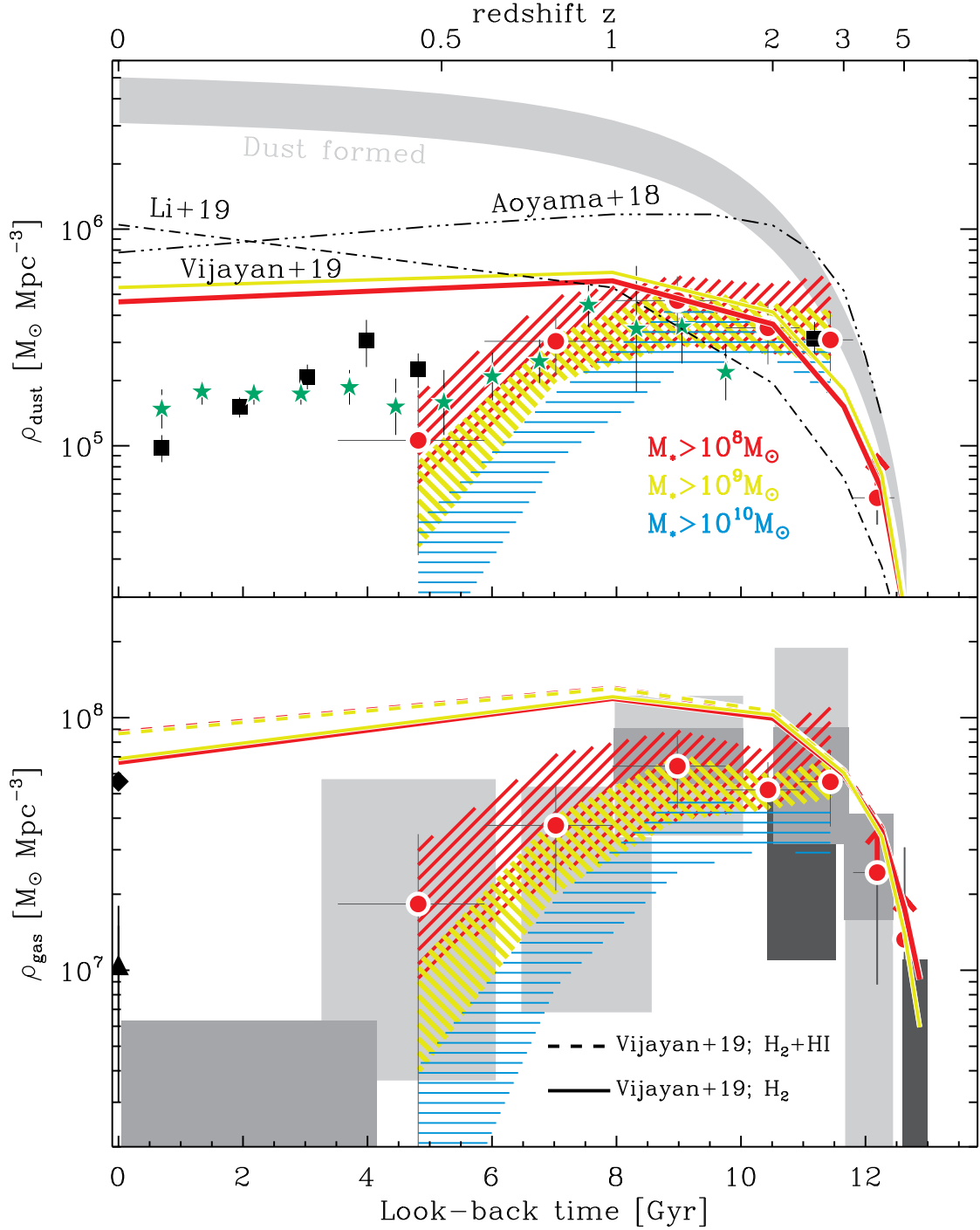}
		\caption{\label{fig:vijayan}
			Same as Figure~\ref{fig:rho_vs_model} but here we compare our measurements to predictions from the semi-analytical model of \citet{vijayan_2019}.
			For $M_{\ast}>10^9\,M_\odot$, we used their `Millennium' predictions, while for $M_{\ast}>10^8\,M_\odot$ we use their `Millennium-II' predictions. 
		}
	\end{center}
	\vspace{0.2cm}
\end{figure*}
\end{document}